\documentclass[10pt] {IEEEtran} 
\pdfoutput=1
\usepackage{booktabs}
\usepackage{cite}
\usepackage[pdftex]{graphicx}
\usepackage[cmex10]{amsmath}
\usepackage{amsfonts} 
\usepackage{array}
\usepackage{mdwmath}
\usepackage{mdwtab}
\usepackage[font=footnotesize, caption=false]{subfig}
\usepackage{fixltx2e}

\usepackage{soul}
\usepackage{verbatim}
\usepackage{url, epstopdf}
\usepackage{color, balance}
\usepackage[ruled]{algorithm2e}
\SetAlgoCaptionSeparator{.}

\makeatletter
\renewcommand*\env@matrix[1][c]{\hskip -\arraycolsep
  \let\@ifnextchar\new@ifnextchar
  \array{*\c@MaxMatrixCols #1}}
\makeatother

\begin{document}

\title{Distributed Widely Linear Frequency Estimation in Unbalanced Three Phase Power Systems}
\author{%
{Sithan Kanna, Dahir H. Dini, Yili Xia, Ron Hui, and Danilo P. Mandic}
\thanks{D. H. Dini, Sithan Kanna and D. P. Mandic are with the Department of Electrical and Electronic Engineering, Imperial College London. Yili Xia is with the School of Information Science and Engineering, Southeast University, China. R. Hui is with Hong Kong University and with Imperial College London. (e-mail: dahir.dini@imperial.ac.uk; ssk08@ic.ac.uk; yili.xia06@gmail.com; d.mandic@imperial.ac.uk; ronhui@eee.hku.hk).}
\thanks{We wish to thank G.K. Supramaniam from TNB Malaysia for his insights and for providing us with the real world measurements.}\vspace{-8mm}
}
%
%
%
%
\maketitle
\begin{abstract}
A novel method for distributed estimation of the frequency of power systems is introduced based on the cooperation between multiple measurement nodes. The proposed distributed widely linear complex Kalman filter (D-ACKF) and the distributed widely linear extended  complex Kalman filter (D-AECKF)  employ a widely linear state space and augmented complex statistics to deal with unbalanced system conditions and the generality complex signals, both second order circular (proper) and second order noncircular (improper). It is shown that the current, strictly linear,  estimators are inadequate for unbalanced systems, a typical case in smart grids, as they do not account for either the noncircularity of Clarke's $\alpha \beta$ voltage in unbalanced conditions or the correlated nature of nodal disturbances.  We illuminate the relationship between the degree of circularity of Clarke's voltage and system imbalance, and  prove that the proposed  widely linear estimators are optimal for such conditions, while also  accounting for the correlated and noncircular nature of real-world nodal disturbances. {Synthetic and real world} case studies over a range of  power system conditions illustrate the theoretical and practical advantages of the proposed methodology. 
\end{abstract}\vspace{-4mm}
%
%
\section{Introduction}
Modern power systems, such as distributed and smart grids, increasingly rely on network-wide information for which signal processing and communication technologies are needed to estimate the power quality parameters. The information  of interest includes system states and operating conditions at different nodes in the system. Modern decentralised and semi-autonomous systems aim to make full use of such information to provide enhanced reliability, efficiency and power quality at both the transmission and distribution side. In addition, the emergence of smart and distributed grids  (e.g. microgrids) has brought to light a number of power quality problems related to the unpredictability of power demand/supply and the accompanying system imbalance {\cite{Bollen_DSP_Power_IEEE_SPM_2009}}.

Power quality refers to the ``fitness'' of electrical power delivered to the consumer, and is characterised by continuity of service, harmonic behaviour, variations of amplitudes of line voltages,  and fluctuations of system frequency around its nominal value {\cite{Bollen_Book}}. Maintaining the system   frequency within its prescribed tolerance range  is a prerequisite  for the health of power systems, as frequency variations are indicative of unbalanced system conditions, such as generation-consumption imbalances or faults. Accurate and robust frequency estimation is therefore a key parameter for both the control and protection of power system and to maintain power quality {\cite{Power_Standard}}.

Currently used frequency estimation techniques include: i) Fourier transform approaches \cite{Wang_DFT_2004,Lobos_Freq_DFT_1997}, ii) gradient decent and least squares adaptive estimation \cite{Xiao_Freq_est_LMS_IEEE_Trans_2005}, and iii) state space methods and Kalman filters \cite{Huang_Freq_est_CEKF_IEEE_Trans_2008,Dash_freq_est_1999,Hu_Kuh_State_MicroGrids_IJCNN_2011,Mangesius_Obradovic_power_grid_2012}.
However, these  are typically designed for  single-phase systems operating under balanced conditions (equal voltage amplitudes) and are not adequate for the demands of modern three--phase and dynamically optimised power systems. 

The modelling of three-phase systems requires simultaneous measurement of the three phase voltages and a mathematical framework to reduce redundancy through transformations of the variables {\cite{Clarke_Inst_Currents_Voltages_1951}}{\cite{Paap_Symmetrical_Components_2000}.  This is typically achieved using Clarke's $\alpha\beta$ transform {\cite{Clarke_Book}}, the output of which is a complex variable $v=v_{\alpha} + \jmath \, v_{\beta}$, that represents balanced systems without loss of information. However, when operating in unbalanced system conditions, standard (strictly linear) complex-valued estimators  inevitably introduce biased estimates together with spurious frequency estimates at twice the system frequency \cite{Beeman_1955}. This is attributed to noncircular phasor trajectories of unbalanced $\alpha\beta$  voltages which  require widely linear estimators to model the system dynamics \cite{ Xia_Adaptive_Frequency_Freq_2012_IEEE_SP_Mag,mandic2014patent, Dini_Three_Phase_Freq_IEEE_IM_2013}.

The existing widely linear frequency estimators are theoretically rigorous but practically restrictive, as they only apply to a single node of a power system. Indeed, distributed and smart grids require cooperation of geographically distributed nodes equipped with local data acquisition and learning capabilities {\cite{Bollen_DSP_Power_IEEE_SPM_2009}}. Distributed estimation and fusion has already found application in both military and civilian scenarios \cite{Stadter_Dist_Spacecraft_IEEEMag_2002,Mandic_Fusion_Book_Short_2008, Cattivelli_Sayed_IEEETranAC_Dist_KF_2010, Olfati_Saber_Flocking_2006}, as cooperation between the nodes (sensors)  provides more accurate and robust estimation over the independent nodes, while approaching the performance of centralised systems at much reduced communication overheads. {Recent approaches include distributed least-mean-square estimation \cite{Lopes_Sayed_TSP_Dist_LMS_2008, Yili_Dist_ACLMS_2011} and Kalman filtering   \cite{Cattivelli_Sayed_IEEETranAC_Dist_KF_2010, Olfati_Saber_Dist_KF, Khan_Dist_KF_IEEETransSP_2008}, however, these references considered models with circular measurement noise without cross-nodal correlations, which is not typical in real world power systems. }

To this end, we introduce a class of distributed sequential state estimators  for the generality of complex signals, both second order circular (proper) and second order noncircular (improper). The state space structure  of the proposed  distributed augmented (widely linear) complex Kalman filter (D-ACKF) and  the distributed augmented extended Kalman filter (D-ACEKF)   also  accounts for the correlation between the observation noises at neighbouring nodes, encountered  when node signals are exposed to  common  sources of interference (harmonics, fluctuations of reacive power). {The aim of this work is therefore  two--fold: (i) to extend existing widely linear frequency estimators {\cite{Dini_Class_WLKF_IEEE_TNNLS_2012,Dini_Three_Phase_Freq_IEEE_IM_2013}} to the distributed scenario; (ii) to investigate the advantages of D-ACKF and D-AECKF over the existing D-CKFs \cite{Olfati_Saber_Dist_KF, Cattivelli_Sayed_IEEETranAC_Dist_KF_2010} through analysis and comprehensive case studies on distributed frequency estimation  under balanced and unbalanced conditions, a key issue in the control and management of microgrids, electricity islands and smart grids.}

\section{Background on widely linear modelling}
Complex-valued signal processing underpins a number of  real-world applications, including wireless communications \cite{Gao_Dist_Cooperative_IEEETranComm_2011}\cite{Mao_Wireless_Comm_IEEETranIFS_2007} and power systems \cite{Xia_WL_Freq_2011}. However, standard approaches are generally suboptimal and are adequate only for a restrictive class of  proper (second order circular) complex processes, for which probability distributions are  rotation invariant \cite{Picinbono97} \cite{Moreno_2008}, \cite{ErikssonAugmentID06}. A zero-mean proper complex signal, $\mathbf{x}$, has equal powers in the real and imaginary part so that the covariance matrix, $\mathbf{R}_{\mathbf{x}}=E\{\mathbf{x}\mathbf{x}^H\}$, is sufficient to represent its complete second-order statistics. However, full statistical description of improper (noncircular) complex signals requires {\em augmented complex statistics} which includes the second order moments called the pseudocovariance, $\mathbf{P}_{\mathbf{x}}=E\{\mathbf{x}\mathbf{x}^T\}$, which models both the power difference and cross-correlation between the real and imaginary parts of a complex signal \cite{Picinbono97}. 

{\bf Widely linear model.} Consider the  minimum mean square error (MSE) estimator of a zero-mean real valued random vector $\mathbf{y}$ in terms of an observed zero-mean real vector $\mathbf{x}$, that is, $\hat{\mathbf{y}} = E\{\mathbf{y}|\mathbf{x}\}$. For jointly normal $\mathbf{y}$ and $\mathbf{x}$, the optimal linear estimator is given by
\begin{equation}
	\hat{\mathbf{y}} = \mathbf{A}\mathbf{x} 
	\label{Eq:Linear_Estimator}
\end{equation}
where $\mathbf{A} =\mathbf{R}_{\mathbf{yx}}\mathbf{R}_{\mathbf{x}}^{-1}$ is a coefficient matrix, and $\mathbf{R}_{\mathbf{yx}} = E\{\mathbf{y}\mathbf{x}^H\}$. Standard, `strictly linear' estimation in ${\mathbb C}$ assumes the same model but with complex valued $\mathbf{y}, \mathbf{x}$, and $\mathbf{A}$. However, when $\mathbf{y}$ and $\mathbf{x}$ are jointly improper $\mathbf{P}_{\mathbf{yx}} = E\{\mathbf{y}\mathbf{x}^T\} \neq \mathbf{0}$, and since $\mathbf{x}$ is also improper $\mathbf{P}_{\mathbf{x}} \neq \mathbf{0}$. The optimal estimator for both proper and improper data is then represented by the widely linear estimator\footnote{The `widely linear' model is associated with the signal generating system, whereas ``augmented statistics'' describe statistical properties of measured signals. Both the terms `widely linear' and `augmented' are used to name the resulting algorithms - in our work we mostly use the term `augmented'.}, given by \cite{Picinbono97}
\begin{equation}
	\hat{\mathbf{y}} = \mathbf{B}\mathbf{x} + \mathbf{C}\mathbf{x}^*  
									 = \mathbf{W} \mathbf{x}^{a}
	\label{Eq:WL_model}
\end{equation}
where $\mathbf{B} = \mathbf{R}_{\mathbf{yx}}\mathbf{D} + \mathbf{P}_{\mathbf{yx}}\mathbf{E}^*$ and $\mathbf{C} = \mathbf{R}_{\mathbf{yx}}\mathbf{E} + \mathbf{P}_{\mathbf{yx}}\mathbf{D}^*$ are coefficient matrices, with $\mathbf{D} = (\mathbf{R}_{\mathbf{x}} - \mathbf{P}_{\mathbf{x}}\mathbf{R}_{\mathbf{x}}^{*-1}\mathbf{P}_{\mathbf{x}}^*)^{-1}$ and $\mathbf{E} = -(\mathbf{R}_{\mathbf{x}} - \mathbf{P}_{\mathbf{x}}\mathbf{R}_{\mathbf{x}}^{*-1}\mathbf{P}_{\mathbf{x}}^*)^{-1}\mathbf{P}_{\mathbf{x}}\mathbf{R}_{\mathbf{x}}^{*-1}$, while $\mathbf{x}^a = [\mathbf{x}^T, \mathbf{x}^H]^T$ is the augmented input vector, and $\mathbf{W} = [\mathbf{B}, \mathbf{C}]$ the optimal coefficient matrix.  Based on ({\ref{Eq:WL_model}}) the full second order statistics for the generality of complex data (proper and improper) is therefore contained in the augmented covariance matrix 
\begin{eqnarray}
\label{rza1}
	\mathbf{R}^a_{\mathbf{x}} = E\{\mathbf{x}^a\mathbf{x}^{aH}\} = \begin{bmatrix} \mathbf{R}_{\mathbf{x}} & \mathbf{P}_{\mathbf{x}}\\\mathbf{P}^*_{\mathbf{x}} & \mathbf{R}^*_{\mathbf{\mathbf{x}}} \end{bmatrix}
\end{eqnarray}
that also includes the pseudocovariance \cite{Picinbono97}\cite{Dini_Class_WLKF_IEEE_TNNLS_2012}\cite{Moreno_2009}.
%
%
%
\section{Diffusion Kalman Filtering}
Every node $i$ in a distributed system (see Figure {\ref{fig:NetwotkTopologyGeneral}}) can be described by a standard linear state space model, given by {\cite{Hayes96}}
\begin{subequations}\label{linearstatespace}
\begin{eqnarray}
	\mathbf{x}_{n} &=& \mathbf{F}_{n-1}\mathbf{x}_{n-1} + \mathbf{w}_{n} \label{s1}\\
	\mathbf{y}_{i,n} &=& \mathbf{H}_{i,n}\mathbf{x}_{n} + \mathbf{v}_{i,n} \label{dm1}
\end{eqnarray}
\end{subequations}
where $\mathbf{x}_{n} \in \mathbb{C}^L$ and $\mathbf{y}_{i,n} \in \mathbb{C}^{K}$ are respectively the state vector at time instant $n$ and observation (measurement) vector at node $i$. The symbol $\mathbf{F}_n$ denotes the state transition matrix, $\mathbf{w}_{n} \in \mathbb{C}^L$  white state noise, while $\mathbf{H}_{i,n}$  is the  observation matrix,   and $\mathbf{v}_{i,n} \in \mathbb{C}^{K}$ is white measurement measurement noise (both at node $i$). Standard state space models assume the noises $\mathbf{w}_{n}$ and $\mathbf{v}_{i,n}$  to be uncorrelated and zero-mean, so that their covariances and pseudocovariance matrices are
\begin{eqnarray}
	E \begin{bmatrix} \mathbf{w}_{n} \\ \mathbf{v}_{i,n} \end{bmatrix} \begin{bmatrix} \mathbf{w}_{k} \\ \mathbf{v}_{i,k} \end{bmatrix}^H
	=
	\begin{bmatrix} \mathbf{Q}_n & \mathbf{0} \\ \mathbf{0} & \mathbf{R}_{i,n} \end{bmatrix} \delta_{nk} \\
	E \begin{bmatrix} \mathbf{w}_{n} \\ \mathbf{v}_{i,n} \end{bmatrix} \begin{bmatrix} \mathbf{w}_{k} \\ \mathbf{v}_{i,k} \end{bmatrix}^T
	=
	\begin{bmatrix} \mathbf{P}_n & \mathbf{0} \\ \mathbf{0} & \mathbf{U}_{i,n} \end{bmatrix} \delta_{nk}
\end{eqnarray}
where $\delta_{nk}$ is the standard Kronecker delta function. 
%
%
\subsection{Distributed Complex Kalman Filter}\label{Sec:D-CKF}
{The distinguishing feature of the proposed class of distributed Kalman filters is that we have used the diffusion strategy in \cite{Cattivelli_Sayed_IEEETranAC_Dist_KF_2010} with more general system and noise models that do not restrict the correlation properties of the cross-nodal observation noises or the signal and noise circularity at different nodes; this generalises existing distributed Kalman filtering algorithms  \cite{Cattivelli_Sayed_IEEETranAC_Dist_KF_2010, Olfati_Saber_Dist_KF, Kar_Moura_Dist_KF_IEEETransSP_2011} to wider application scenarios.} Figure \ref{fig:NetwotkTopologyGeneral} illustrates the distributed estimation scenario; the highlighted  neighbourhood of node $i$ compromises the set of nodes, denoted by $\mathcal{N}_i$, that communicate  with the node $i$ (including Node $i$ itself).  The state estimate at node $i$ with a complex Kalman filter (CKF)  is then based on all the data from the neighbourhood $\mathcal{N}_i$ consisting of $M = |\mathcal{N}_i|$ nodes, and is denoted by $\mathbf{\widehat{\underline{x}}}_{i,n|n}$, where the symbol $|\mathcal{N}_i|$ denotes the number of nodes in the neighbourhood $\mathcal{N}_i$. Finally, the collective neighbourhood observation equation at node $i$ is given by
\begin{eqnarray} \label{c_neighbourhood_obs_eqn}
	\mathbf{\underline{y}}_{i,n} = \mathbf{\underline{H}}_{i,n}\mathbf{x}_{n} + \mathbf{\underline{v}}_{i,n}
\end{eqnarray}
while the collective (neighbourhood) variables are defined as 
\begin{align*}
	\mathbf{\underline{y}}_{i,n} &= \begin{bmatrix} \mathbf{y}_{i_1,n}^T, \mathbf{y}_{i_2,n}^T, \ldots , \mathbf{y}_{i_M,n}^T \end{bmatrix}^T\\ 
	\mathbf{\underline{H}}_{i,n} &= \begin{bmatrix} \mathbf{H}_{i_1,n}^T, \mathbf{H}_{i_2,n}^T, \ldots , \mathbf{H}_{i_M,n}^T \end{bmatrix}^T\\
	\mathbf{\underline{v}}_{i,n} &= \begin{bmatrix} \mathbf{v}_{i_1,n}^T, \mathbf{v}_{i_2,n}^T, \ldots , \mathbf{v}_{i_M,n}^T \end{bmatrix}^T
\end{align*}
where $\{i_1, i_2, \ldots, i_M\}$ are the nodes in the neighbourhood $\mathcal{N}_i$. The covariance and pseudocovariance matrices of the collective observation noise vector then become
\begin{align*}
	&\mathbf{\underline{R}}_{i,n} = E\{\mathbf{\underline{v}}_{i,n}\mathbf{\underline{v}}_{i,n}^H \} 
	= 
	\begin{bmatrix}[l] \mathbf{R}_{i_1,n}    	& \mathbf{R}_{i_1i_2,n} & \cdots & \mathbf{R}_{i_1i_M,n}  \\
	    						\mathbf{R}_{i_2i_1,n} 	& \mathbf{R}_{i_2,n}    & \cdots & \mathbf{R}_{i_2i_M,n}  \\
	    						\vdots 									& \vdots 								& \ddots & \vdots 								\\
	    						\mathbf{R}_{i_Mi_1,n}  	& \mathbf{R}_{i_Mi_2,n} & \cdots & \mathbf{R}_{i_M,n}				
	\end{bmatrix} \\
	&\mathbf{\underline{U}}_{i,n} = E\{\mathbf{\underline{v}}_{i,n}\mathbf{\underline{v}}_{i,n}^T \}
	= 
	\begin{bmatrix}[l] \mathbf{U}_{i_1,n}    	& \mathbf{U}_{i_1i_2,n} & \cdots & \mathbf{U}_{i_1i_M,n}  \\
	    						\mathbf{U}_{i_2i_1,n} 	& \mathbf{U}_{i_2,n}    & \cdots & \mathbf{U}_{i_2i_M,n}  \\
	    						\vdots 									& \vdots 								& \ddots & \vdots 								\\
	    						\mathbf{U}_{i_Mi_1,n}  	& \mathbf{U}_{i_Mi_2,n} & \cdots & \mathbf{U}_{i_M,n}				
	\end{bmatrix} 	
\end{align*}
where $\mathbf{R}_{i_a,n} = E\{\mathbf{v}_{i_a,n}\mathbf{v}_{i_a,n}^H\}$, $\mathbf{R}_{i_ai_b,n} = E\{\mathbf{v}_{i_a,n}\mathbf{v}_{i_b,n}^H\}$, $\mathbf{U}_{i_a,n} = E\{\mathbf{v}_{i_a,n}\mathbf{v}_{i_a,n}^T\}$ and $\mathbf{U}_{i_ai_b,n} = E\{\mathbf{v}_{i_a,n}\mathbf{v}_{i_b,n}^T\}$, for $a,b \in \{1,2,\ldots , M\}$. 
\vspace{3mm}\\
\noindent
{\bf Diffusion step.} The so found local neighbourhood state estimates are followed by the diffusion  (combination) step, 
%
\begin{figure}[t]
  \centering  
  \includegraphics[trim = 4mm 30mm 4mm 9mm,scale=0.8]{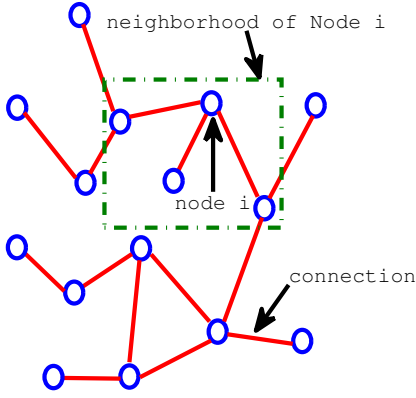}
  \caption{An  example of a distributed network topology.}
  \label{fig:NetwotkTopologyGeneral}
\end{figure}
%
\begin{eqnarray}
	\mathbf{\widehat{x}}_{i,n|n} = \sum_{k \in \mathcal{N}_i} c_{k,i}\mathbf{\widehat{\underline{x}}}_{k,n|n} \label{eq:diff_step}
\end{eqnarray}
which calculates the diffused state estimates $\mathbf{\widehat{x}}_{i,n|n}$ as a weighted sum of the estimates from the neighbourhood $\mathcal{N}_i$, where $c_{k,i} \geq 0$ are the weighting coefficients satisfying $\sum_{k \in \mathcal{N}_i} c_{k,i} = 1$. 
{The combination weights $c_{k,i}$ used by the diffusion step in \eqref{eq:diff_step} can follow the Metropolis \cite{Lopes_Sayed_TSP_Dist_LMS_2008}, Laplacian \cite{Bru_Convergence_1994} or the nearest neighbour \cite{Cattivelli_Sayed_IEEETranAC_Dist_KF_2010} rules. However, finding the set of optimal weights is a challenging task, though progress has been made in important cases\cite{Li_Optimal_Dist_fusion_2003, Sayed_Proceedings_2014, Sayed_arXiv}.} 

The distributed complex Kalman filter (D-CKF) aims to approach the performance of a centralised Kalman filter (access to  data from all the nodes) via neighbourhood collaborations and diffusion, and is summarised in Algorithm \ref{alg:D-CKF}. Each node within D-CKF forms a collective observation as in \eqref{c_neighbourhood_obs_eqn}, using information from its neighbours; thereafter, each node computes a neighbourhood state estimate which is again shared with neighbours in order to be used for the diffusion step. 
%
\begin{algorithm}[t]
Initialisation: For each node $i=1,2,\ldots,N$
\begin{eqnarray}
	\mathbf{\widehat{x}}_{i,0|0} &=& E\{\mathbf{x}_{0}\} \nonumber\\
	\mathbf{M}_{i,0|0} &=& E\{(\mathbf{x}_{0}-E\{\mathbf{x}_{0}\})(\mathbf{x}_{0}-E\{\mathbf{x}_{0}\})^H\} \nonumber
\end{eqnarray} 
For every time instant $n=1,2,\ldots$ 

$-$ Evaluate at each node $i=1,2,\ldots,N$
\begin{align}
	\mathbf{\widehat{x}}_{i,n|n-1} &= \mathbf{F}_{n-1}\mathbf{\widehat{x}}_{i,n-1|n-1} \label{px_est}\\
  \mathbf{M}_{i,n|n-1} &= \mathbf{F}_{n-1}\mathbf{M}_{i,n-1|n-1}\mathbf{F}^{H}_{n-1} + \mathbf{Q}_n \label{pMSE}\\
	\mathbf{G}_{i,n} &= \mathbf{M}_{i,n|n-1}\mathbf{\underline{H}}^{H}_{i,n}\big(\mathbf{\underline{H}}_{i,n}\mathbf{M}_{i,n|n-1}\mathbf{\underline{H}}^{H}_{i,n} + \mathbf{\underline{R}}_{i,n}\big)^{-1} \label{gain}\\
	\mathbf{\widehat{\underline{x}}}_{i,n|n} &= \mathbf{\widehat{x}}_{i,n|n-1} + \mathbf{G}_{i,n}\big(\mathbf{\underline{y}}_{i,n} - \mathbf{\underline{H}}_{i,n}\mathbf{\widehat{x}}_{i,n|n-1}\big)\label{x_est}\\
	\mathbf{M}_{i,n|n} &= (\mathbf{I} - \mathbf{G}_{i,n}\mathbf{\underline{H}}_{i,n})\mathbf{M}_{i,n|n-1}\label{MSE}
\end{align}
$-$ For every node $i$, compute the diffusion update as
\begin{align}\label{Diff}
	\mathbf{\widehat{x}}_{i,n|n} = \sum_{k \in \mathcal{N}_i} c_{k,i}\mathbf{\widehat{\underline{x}}}_{k,n|n}
\end{align}
\caption{The D-CKF}
\label{alg:D-CKF}
\end{algorithm}
\\ \vspace{1mm}

\textbf{Remark \#1:} The D-CKF algorithm\footnote{The matrices $\mathbf{M}_{i,n|n}$ and $\mathbf{M}_{i,n|n-1}$ do not represent the covariances of $\mathbf{\widehat{x}}_{i,n|n}$ and $\mathbf{\widehat{x}}_{i,n|n-1}$, as is the case for the standard Kalman filter operating on linear Gaussian systems. This is due to the use of the suboptimal diffusion step, which updates the state estimate and not the covariance matrix $\mathbf{M}_{i,n|n}$.} given in Algorithm \ref{alg:D-CKF} is a variant of \cite{Cattivelli_Sayed_IEEETranAC_Dist_KF_2010}. It employs the standard (strictly linear) state space model \eqref{linearstatespace}, and therefore does not cater for widely linear complex state space models or noncircular state and observation noises ($\mathbf{P}_{n} \neq \mathbf{0}$ and $\mathbf{U}_{i,n} \neq \mathbf{0}$ for $i=1,\ldots,N$). 
{Unlike existing distributed complex Kalman filters, the proposed D-CKF algorithm in \eqref{px_est} -- \eqref{Diff}  caters for the correlations between the neighbourhood observation noises. When no such correlations exits, is identical to Algorithm \ref{alg:D-CKF} in \cite{Cattivelli_Sayed_IEEETranAC_Dist_KF_2010}.}
%
%
\subsection{Distributed Augmented Complex Kalman Filter} \label{Sec:D-ACKF}
As stated in Remark \#1, current state space algorithms do not cater for  widely linear state and observation models or for improper measurements, states, and state and observation noises. To this end,  we next  employ the widely linear model in \eqref{Eq:WL_model} to introduce the widely linear version of the standard, strictly linear, distributed state space model\footnote{Observe that the noise models can also be widely linear, in which case:\\ $\mathbf{w}_{n} =\mathbf{D}_n\acute{\mathbf{w}}_{n} + \mathbf{E}_n\acute{\mathbf{w}}^*_{n}\quad$ and $\quad\mathbf{v}_{n} =\mathbf{F}_n\acute{\mathbf{v}}_{n} + \mathbf{H}_n\acute{\mathbf{v}}^*_{n}$, where $\mathbf{D, E, F, H}$ are coefficient matrices and $\acute{\mathbf{w}}_{n}$ and $\acute{\mathbf{v}}_{n}$ are proper or improper noise models.}   in \eqref{linearstatespace} (see also \cite{Dini_Class_WLKF_IEEE_TNNLS_2012}, \cite{Dini_Dist_ACKF_ASILOMAR_2012})
\begin{subequations}\label{WLstatespace}
\begin{eqnarray}
	\mathbf{x}_{n} &=& \mathbf{F}_{n-1}\mathbf{x}_{n-1} + \mathbf{A}_{n-1}\mathbf{x}^*_{n-1} + \mathbf{w}_{n} \\
	\mathbf{y}_{i,n} &=& \mathbf{H}_{i,n}\mathbf{x}_{n} + \mathbf{B}_{i,n}\mathbf{x}_{n}^* + \mathbf{v}_{i,n} \label{WLdm1}
\end{eqnarray}  
\end{subequations}
The compact, augmented representation, of this model is 
\begin{subequations}\label{AUGstatespace}
\begin{eqnarray}
	\mathbf{x}^a_{n} &=& \mathbf{F}^a_{n-1}\mathbf{x}^a_{n-1} + \mathbf{w}^a_{n} \label{as1}\\
	\mathbf{y}_{i,n}^a &=& \mathbf{H}_{i,n}^a\mathbf{x}_{n}^a + \mathbf{v}_{i,n}^a 
\end{eqnarray} 
\end{subequations}  
where $\mathbf{x}^a_n = [\mathbf{x}^T_n, \mathbf{x}^H_n]^T$ and $\mathbf{y}^a_n = [\mathbf{y}^T_n, \mathbf{y}^H_n]^T$, while
\begin{eqnarray*}
	\mathbf{F}^a_{n} = \begin{bmatrix} \mathbf{F}_{n} & \mathbf{A}_{n}\\ \mathbf{A}^*_{n} & \mathbf{F}^*_{n} \end{bmatrix}
	\text{  and  }
	\mathbf{H}^a_{i,n} = \begin{bmatrix} \mathbf{H}_{i,n} & \mathbf{B}_{i,n}\\ \mathbf{B}^*_{i,n} & \mathbf{H}^*_{i,n} \end{bmatrix}
\end{eqnarray*} 			
For \textit{strictly linear systems}, $\mathbf{A}_{n} = \mathbf{0}$ and $\mathbf{B}_{i,n} = \mathbf{0}$, so that the widely linear (augmented) state space model degenerates into a strictly linear one. However, the augmented state space representation above is still preferred in order to account for the pseudocovariances which reflect the \textit{impropriety of the signals} (\textit{cf.} widely linear nature of systems). 

The augmented covariance matrices of $\mathbf{w}^a_n = [\mathbf{x}^T_n, \mathbf{w}^H_n]^T$ and $\mathbf{v}^a_{i,n} = [\mathbf{v}^T_{i,n}, \mathbf{v}^H_{i,n}]^T$ are then given by
\begin{eqnarray}
	\mathbf{Q}^a_n &=& E\{\mathbf{w}^a_{n}\mathbf{w}^{aH}_{n}\} =\begin{bmatrix} \mathbf{Q}_n & \mathbf{P}_n\\ \mathbf{P}^*_n & \mathbf{Q}^*_n \end{bmatrix} \label{aQ}\\
	\mathbf{R}^a_{i,n} &=& E\{\mathbf{v}^a_{i,n}\mathbf{v}^{aH}_{i,n}\} =\begin{bmatrix} \mathbf{R}_{i,n} & \mathbf{U}_{i,n}\\ \mathbf{U}^*_{i,n} & \mathbf{R}^*_{i,n} \end{bmatrix}\label{aR}	
\end{eqnarray}
\noindent
{\bf Neighbourhood variables.} For collaborative estimation of the state within distributed networks,  neighbourhood observation equations use all available neighbourhood observation data, to give
\begin{eqnarray} \label{ac_neighbourhood_obs_eqn}
	\mathbf{\underline{y}}_{i,n} = \mathbf{\underline{H}}_{i,n}\mathbf{x}_{n} + \mathbf{\underline{B}}_{i,n}\mathbf{x}_{n}^* +\mathbf{\underline{v}}_{i,n}
\end{eqnarray}
where the symbol $\mathbf{\underline{B}}_{i,n} = \big[\mathbf{B}_{i_1,n}^T, \mathbf{B}_{i_2,n}^T, \ldots, \mathbf{B}_{i_M,n}^T \big]^T$ denotes the conjugate state matrix , and $\{i_1, i_2, \ldots, i_M\} \in \mathcal{N}_i$. The augmented neighbourhood observation equations now become
\begin{eqnarray}\label{aug_y_collective}
	\mathbf{\underline{y}}_{i,n}^a = \mathbf{\underline{H}}_{i,n}^a\mathbf{x}_{n}^a + \mathbf{\underline{v}}_{i,n}^a 
\end{eqnarray}
with the augmented neighbourhood variables defined as
\begin{eqnarray}
	\mathbf{\underline{y}}_{i,n}^a = \begin{bmatrix} \mathbf{\underline{y}}_{i,n} \\ \mathbf{\underline{y}}_{i,n}^* \end{bmatrix}, 
	\quad
	\mathbf{\underline{H}}_{i,n}^a = \begin{bmatrix} \mathbf{\underline{H}}_{i,n} & \mathbf{\underline{B}}_{i,n}\\ 
																	 \mathbf{\underline{B}}_{i,n}^* & \mathbf{\underline{H}}_{i,n}^*\end{bmatrix},
	\quad
	\mathbf{\underline{v}}_{i,n}^a = \begin{bmatrix} \mathbf{\underline{v}}_{i,n} \\ \mathbf{\underline{v}}_{i,n}^* \end{bmatrix}
\end{eqnarray}
Consequently, the covariance of the augmented neigbourhood observation noise $\mathbf{\underline{v}}_{i,n}^a$ takes the form
\begin{eqnarray}\label{aug_R_collective}
	\mathbf{\underline{R}}_{i,n}^a 
	= E\{\mathbf{\underline{v}}^a_{i,n}\mathbf{\underline{v}}^{aH}_{i,n}\} 
	= \begin{bmatrix} \mathbf{\underline{R}}_{i,n} & \mathbf{\underline{U}}_{i,n}\\ \mathbf{\underline{U}}^*_{i,n} & \mathbf{\underline{R}}^*_{i,n} \end{bmatrix}	
\end{eqnarray}
{\textbf{Remark \#2:} Observe that \eqref{aug_R_collective} caters for both the covariances $E\{\mathbf{v}_{i,n}\mathbf{v}_{i,n}^H\}$ and cross-correlations $E\{\mathbf{v}_{i,n}\mathbf{v}_{k,n}^H\}$, $i\neq k$ between the nodal observation noises. This is achieved through the covariance matrix $\mathbf{\underline{R}}_{i,n}$ and the pseudocovariances $E\{\mathbf{v}_{i,n}\mathbf{v}_{i,n}^T\}$, while the  cross-pseudocorrelations $E\{\mathbf{v}_{i,n}\mathbf{v}_{k,n}^T\}$ are accounted for through the pseudocovariance matrix $\mathbf{\underline{U}}_{i,n}$.}

Finally, the augmented diffused state estimate becomes
\begin{eqnarray}
	\mathbf{\widehat{x}}_{i,n|n}^a = \sum_{k \in \mathcal{N}_i} c_{k,i}\mathbf{\widehat{\underline{x}}}_{k,n|n}^a
\end{eqnarray}
and represents a weighted average of the augmented (neighbourhood) state estimates. The proposed distributed augmented complex Kalman filter (D-ACKF), based on the widely linear state space model, is summarised in Algorithm \ref{alg:D-ACKF}. 

For strictly linear systems ($\mathbf{A}_{n} = \mathbf{0}$ and $\mathbf{B}_{i,n} = \mathbf{0}$ for all $n$ and $i$) and in the presence of circular state and observation noises ($\mathbf{P}_{n} = \mathbf{0}$ and $\mathbf{U}_{i,n} = \mathbf{0}$ for all $n$ and $i$), the D-ACKF and D-CKF algorithms yield identical state estimates for all time instants $n$.   However, the D-ACKF is more general than the D-CKF, since it  also caters for the noncircular data and noise natures together with correlated state and observation noises.
\begin{algorithm}[t]
Initialisation: For each node $i=1,2,\ldots,N$
\begin{eqnarray*}
	\mathbf{\widehat{x}}_{i,0|0}^a &=& \big[E\{\mathbf{x}_{0}\}^T, E\{\mathbf{x}_{0}\}^H\big]^T \\
	\mathbf{M}_{i,0|0}^a &=& E\big\{(\mathbf{x}_{0}^a-\mathbf{\widehat{x}}_{i,0|0}^a)(\mathbf{x}_{0}^a-\mathbf{\widehat{x}}_{i,0|0}^a)^{aH}\big\}
\end{eqnarray*} 
For every time instant $n=1,2,\ldots$ 

\hspace{0.15cm}$-$ Evaluate at each node $i=1,2,\ldots,N$
\begin{align}
	\mathbf{\widehat{x}}_{i,n|n-1}^a &= \mathbf{F}_{n-1}^a\mathbf{\widehat{x}}_{i,n-1|n-1}^a \label{augpx_est}\\
  \mathbf{M}_{i,n|n-1}^a &= \mathbf{F}_{n-1}^a\mathbf{M}_{i,n-1|n-1}^a\mathbf{F}^{aH}_{n-1} + \mathbf{Q}_n^a \label{augpMSE}\\
	\mathbf{G}_{i,n}^a &= \mathbf{M}_{i,n|n-1}^a\mathbf{\underline{H}}^{aH}_{i,n}\big(\mathbf{\underline{H}}_{i,n}^a\mathbf{M}_{i,n|n-1}^a\mathbf{\underline{H}}^{aH}_{i,n} + \mathbf{\underline{R}}_{i,n}^a\big)^{-1} \label{aug_gain}\\
	\mathbf{\widehat{\underline{x}}}_{i,n|n}^a &= \mathbf{\widehat{x}}_{i,n|n-1}^a + \mathbf{G}_{i,n}^a\big(\mathbf{\underline{y}}_{i,n}^a - \mathbf{\underline{H}}_{i,n}^a\mathbf{\widehat{x}}_{i,n|n-1}^a\big)\label{augx_est}\\
	\mathbf{M}_{i,n|n}^a &= (\mathbf{I} - \mathbf{G}_{i,n}^a\mathbf{\underline{H}}_{i,n}^a)\mathbf{M}_{i,n|n-1}^a\label{augMSE}
\end{align}
\hspace{0.15cm}$-$ For every node $i$, compute the diffusion update as
\begin{align}
	\mathbf{\widehat{x}}_{i,n|n}^a = {\sum}_{k \in \mathcal{N}_i} c_{k,i}\mathbf{\widehat{\underline{x}}}_{k,n|n}^a \label{augdx}
\end{align}
\caption{The D-ACKF}
\label{alg:D-ACKF}
\end{algorithm}
\newline
\textbf{Remark \#3:} The information form of the D-ACKF, given in Algorithm \ref{alg:D-ACKF_info}, can be used to cater for the noncircularity of data when  observation noises at different nodes are uncorrelated. Moreover,  nodes in the distributed network can switch between the general D-ACKF in Algorithm \ref{alg:D-ACKF} and the information form of D-ACKF in Algorithm \ref{alg:D-ACKF_info}, depending on the correlation between the observation noises.
\begin{algorithm}[t]
Initialisation: For each node $i=1,2,\ldots,N$
\begin{eqnarray*}
	\mathbf{\widehat{x}}_{i,0|0}^a &=& \big[E\{\mathbf{x}_{0}\}^T, E\{\mathbf{x}_{0}\}^H\big]^T \\
	\mathbf{M}_{i,0|0}^a &=& E\big\{(\mathbf{x}_{0}^a-\mathbf{\widehat{x}}_{i,0|0}^a)(\mathbf{x}_{0}^a-\mathbf{\widehat{x}}_{i,0|0}^a)^{aH}\big\}
\end{eqnarray*} 
For every time instant $n=1,2,\ldots$ 

\hspace{0.15cm}$-$ Evaluate at each node $i=1,2,\ldots,N$
\begin{align}
	\mathbf{\widehat{x}}_{i,n|n-1}^a &= \mathbf{F}_{n-1}^a\mathbf{\widehat{x}}_{i,n-1|n-1}^a \label{augpx_est_info}\\
  \mathbf{M}_{i,n|n-1}^a &= \mathbf{F}_{n-1}^a\mathbf{M}_{i,n-1|n-1}^a\mathbf{F}^{aH}_{n-1} + \mathbf{Q}_n^a \label{augpMSE_info}\\
	\mathbf{S}_{i,n}^a &= {\sum}_{k \in \mathcal{N}_i} \mathbf{H}^{aH}_{k,n}(\mathbf{R}_{k,n}^a)^{-1}\mathbf{H}_{k,n}^a \\
	\mathbf{r}_{i,n}^a &= {\sum}_{k \in \mathcal{N}_i} \mathbf{H}^{aH}_{k,n}(\mathbf{R}_{k,n}^a)^{-1}\mathbf{y}_{k,n}^a \\
	(\mathbf{M}_{i,n|n}^a)^{-1} &= (\mathbf{M}_{i,n|n-1}^a)^{-1} + \mathbf{S}_{i,n}^a\\
	\mathbf{\widehat{\chi}}_{i,n|n}^a &= \mathbf{\widehat{x}}_{i,n|n-1}^a + \mathbf{M}_{i,n|n}^a\big(\mathbf{r}_{i,n}^a - \mathbf{S}_{i,n}^a\mathbf{\widehat{x}}_{i,n|n-1}^a\big)
\end{align}
\hspace{0.15cm}$-$ For every node $i$, compute the diffusion update as
\begin{align}
	\mathbf{\widehat{x}}_{i,n|n}^a = {\sum}_{k \in \mathcal{N}_i} c_{k,i}\mathbf{\widehat{\chi}}_{i,n|n}^a \label{augdx_info}
\end{align}
\caption{The D-ACKF Information Form}
\label{alg:D-ACKF_info}
\end{algorithm}
%
%
%
\subsection{Bias Analysis of the D-ACKF Estimates}
%
%
Consider the following augmented complex variables:  the local (non-diffused) error at node $i \in [1,N]$ given by $\mathbf{\underline{e}}_{i,n|n}^a = \mathbf{x}_{n}^a - \mathbf{\widehat{\underline{x}}}_{i,n|n}^a$,  the prediction error $\mathbf{e}_{i,n|n-1}^a = \mathbf{x}_{n}^a - \mathbf{\widehat{x}}_{i,n|n-1}^a$, and the diffused error $\mathbf{e}_{i,n|n}^a = \mathbf{x}_{n}^a - \mathbf{\widehat{x}}_{i,n|n}^a$. 
Then%
\begin{align}\label{MeanaugDiffError2}
	E\{\mathbf{e}_{i,n|n}^a\}	&= \sum_{k \in \mathcal{N}_i} c_{k,i}
	\mathbf{M}_{k,n|n}^a(\mathbf{M}_{k,n|n-1}^a)^{-1}\mathbf{F}_{n-1}^aE\{\mathbf{e}_{k,n-1|n-1}^a\} \nonumber\\
	&= \mathbf{0}
\end{align}
\noindent
\textbf{Remark \#4:} The expression \eqref{MeanaugDiffError2} shows that the D-ACKF is an unbiased estimator of  both proper and improper complex random signals.
%
%
%
%
\section{Distributed Augmented Complex Extended Kalman Filter} \label{Sec:D-ACEKF}
We next introduce the distributed augmented complex extended Kalman filter (D-ACEKF) for  nonlinear state space models of the form  
\begin{subequations}\label{nonlinearSS}
\begin{eqnarray}
	\mathbf{x}_n &=& \mathbf{f}[\mathbf{x}_{n-1}] + \mathbf{w}_n \label{nsm1} \\
	\mathbf{y}_{i,n} &=& \mathbf{h}_i[\mathbf{x}_n] + \mathbf{v}_{i,n}  \label{nom1}
\end{eqnarray}
\end{subequations}
where the nonlinear functions $\mathbf{f}[\cdot]$ and $\mathbf{h}_i[\cdot]$ are respectively the (possibly time varying) process model and observation model at node $i$, the  remaining variables are as defined above. Within the extended Kalman filter (EKF) framework, the nonlinear state and observation functions are approximated by their first order Taylor series expansions (TSE) about the state estimates $\mathbf{\widehat{x}}_{i,n-1|n-1}$ and $\mathbf{\widehat{x}}_{i,n|n-1}$ for each node $i$, so that  {\cite{Dini_ACEKF_UDRC_2011}}
\begin{subequations}\label{linearisedSS}
\begin{align}
	\mathbf{x}_n &\approx \mathbf{F}_{i,n-1}\mathbf{x}_{n-1} + \mathbf{A}_{i,n-1}\mathbf{x}^*_{n-1} + \mathbf{r}_{i,n-1} + \mathbf{w}_n \label{eqn-TSEstate}\\ 
	\mathbf{y}_{i,n} &\approx \mathbf{H}_{i,n}\mathbf{x}_{n} + \mathbf{B}_{i,n}\mathbf{x}^*_{n} + \mathbf{z}_{i,n} + \mathbf{v}_{i,n} \label{eqn-TSEobs}
\end{align}
\end{subequations}
where the Jacobians of functions  $\mathbf{f}[\cdot]$ and $\mathbf{h}_i[\cdot]$ are defined as
\begin{align}
	\mathbf{F}_{i,n} &= \frac{\partial \mathbf{f}[\mathbf{x}]}{\partial\mathbf{x}}\Big|_{\mathbf{x} = \mathbf{\widehat{x}}_{i,n|n}} 
	\hspace{0.1cm}\text{,}\hspace{0.3cm}\qquad
	\mathbf{A}_{i,n} = \frac{\partial \mathbf{f}[\mathbf{x}]}{\partial\mathbf{x}^*}\Big|_{\mathbf{x}^* = \mathbf{\widehat{x}}^*_{i,n|n}} 
	\text{,}\nonumber\\
	\mathbf{H}_{i,n} &= \frac{\partial \mathbf{h}_i[\mathbf{x}]}{\partial\mathbf{x}}\Big|_{\mathbf{x} = \mathbf{\widehat{x}}_{i,n|n-1}} 
	\hspace{0.1cm}\text{and}\hspace{0.2cm}
	\mathbf{B}_{i,n} = \frac{\partial \mathbf{h}_i[\mathbf{x}]}{\partial\mathbf{x}^*}\Big|_{\mathbf{x}^* = \mathbf{\widehat{x}}^*_{i,n|n-1}} \nonumber   
\end{align}
and the vectors 
\begin{align*}
	\mathbf{r}_{i,n} &= \mathbf{f}[\mathbf{\widehat{x}}_{i,n-1|n-1}] - \mathbf{F}_{i,n-1}\mathbf{\widehat{x}}_{i,n-1|n-1} - \mathbf{A}_{i,n-1}\mathbf{\widehat{x}}^*_{i,n-1|n-1}	\\
	\mathbf{z}_{i,n} &= \mathbf{h}_i[\mathbf{\widehat{x}}_{i,n|n-1}] - \mathbf{H}_{i,n}\mathbf{\widehat{x}}_{i,n|n-1} - \mathbf{B}_{i,n}\mathbf{\widehat{x}}^*_{i,n|n-1}
\end{align*}
are deterministic inputs calculated from the state space model and state estimate. In order to cater for the full second order statistics of the variables in the linearised state space  in \eqref{linearisedSS}, we shall consider its compact (augmented) version   given by
\begin{subequations}\label{augLinearisedSS}
\begin{align}
	\mathbf{x}^a_n &\approx  \mathbf{F}^a_{i,n-1}\mathbf{x}^a_{n-1} + \mathbf{r}^a_{i,n-1} + \mathbf{w}^a_n \\ 
	\mathbf{y}^a_{i,n} &\approx  \mathbf{H}^a_{i,n}\mathbf{x}^a_{n} + \mathbf{z}^a_{i,n} + \mathbf{v}^a_{i,n} 
\end{align} 
\end{subequations}
where $\mathbf{r}^a_{i,n} = \big[ \mathbf{r}^T_{i,n} , \mathbf{r}^H_{i,n} \big]^T$, $\mathbf{z}^a_{i,n} = \big[ \mathbf{z}^T_{i,n} , \mathbf{z}^H_{i,n} \big]^T$, while\\\\ 			
			$\mathbf{F}^a_{i,n} = \begin{bmatrix} \mathbf{F}_{i,n} & \mathbf{A}_{i,n}\\ \mathbf{A}^*_{i,n} & \mathbf{F}^*_{i,n} \end{bmatrix}$ \quad and \quad 
			$\mathbf{H}^a_{i,n} = \begin{bmatrix} \mathbf{H}_{i,n} & \mathbf{B}_{i,n}\\ \mathbf{B}^*_{i,n} & \mathbf{H}^*_{i,n} \end{bmatrix}$.\\\\

Observe that the collective neighbourhood augmented observation equation for node $i$ takes the form
\begin{eqnarray} \label{neighbourhood_obs_eqn_NL}
	\mathbf{\underline{y}}^a_{i,n} = \mathbf{\underline{h}}_i^a[\mathbf{x}_{n}] + \mathbf{\underline{v}}^a_{i,n}
\end{eqnarray}
while the collective observation function defined as
\begin{eqnarray*}
	\mathbf{\underline{h}}_i^a[\mathbf{x}_{n}] &=& \Big[\mathbf{\underline{h}}^T_i[\mathbf{x}_{n}], \mathbf{\underline{h}}^H_i[\mathbf{x}_{n}]\Big]^T \\
	\mathbf{\underline{h}}_i[\mathbf{x}_{n}] &=& \Big[\mathbf{h}^T_{i_1}[\mathbf{x}_{n}], \mathbf{h}^T_{i_2}[\mathbf{x}_{n}], \ldots, 
																										\mathbf{h}^T_{i_M}[\mathbf{x}_{n}]\Big]^T
\end{eqnarray*}
where $i \in \{i_1, i_2, \ldots, i_M\}$ are all the nodes in the neighbourhood $\mathcal{N}_i$. The first order approximation of \eqref{neighbourhood_obs_eqn_NL} is then 
\begin{align}
	\mathbf{\underline{y}}^a_{i,n} &\approx  \mathbf{\underline{H}}^a_{i,n}\mathbf{x}^a_{n} + \mathbf{\underline{z}}^a_{i,n} + \mathbf{\underline{v}}^a_{i,n} 
\end{align}
with the Jacobian of the collective observation function 
\begin{eqnarray*}
	\mathbf{\underline{H}}^a_{i,n} = \begin{bmatrix} \mathbf{\underline{H}}_{i,n} & \mathbf{\underline{B}}_{i,n}\\ \mathbf{\underline{B}}^*_{i,n} & \mathbf{\underline{H}}^*_{i,n} \end{bmatrix}
\end{eqnarray*}
where $\mathbf{\underline{H}}_{i,n} = \big[\mathbf{H}_{i_1,n}^T, \mathbf{H}_{i_2,n}^T, \ldots, \mathbf{H}_{i_M,n}^T \big]^T$ and
$\mathbf{\underline{B}}_{i,n} = \big[\mathbf{B}_{i_1,n}^T, \mathbf{B}_{i_2,n}^T, \ldots, \mathbf{B}_{i_M,n}^T \big]^T$, wherein 
\begin{align}
	\mathbf{H}_{i_k,n} &= \frac{\partial \mathbf{h}_{i_k}[\mathbf{x}]}{\partial\mathbf{x}}\Big|_{\mathbf{x} = \mathbf{\widehat{x}}_{i,n|n-1}} 
	\hspace{0.1cm}\text{and}\hspace{0.2cm}
	\mathbf{B}_{i_k,n} = \frac{\partial \mathbf{h}_{i_k}[\mathbf{x}]}{\partial\mathbf{x}^*}\Big|_{\mathbf{x}^* = \mathbf{\widehat{x}}^*_{i,n|n-1}} \nonumber   
\end{align}
%
%
\begin{algorithm}[t]
Initialisation: For each node $i=1,2,\ldots,N$
\begin{eqnarray*}
	\mathbf{\widehat{x}}_{i,0|0}^a &=& \big[E\{\mathbf{x}_{0}\}^T, E\{\mathbf{x}_{0}\}^H\big]^T \\
	\mathbf{M}_{i,0|0}^a &=& E\big\{(\mathbf{x}_{0}^a-\mathbf{\widehat{x}}_{i,0|0}^a)(\mathbf{x}_{0}^a-\mathbf{\widehat{x}}_{i,0|0}^a)^{aH}\big\}
\end{eqnarray*} 
For every time instant $n=1,2,\ldots$ 

\hspace{0.15cm}$-$ Evaluate at each node $i=1,2,\ldots,N$
\begin{align}
	\mathbf{\widehat{x}}_{i,n|n-1}^a &= \Big[\mathbf{f}^T[\mathbf{\widehat{x}}_{i,n-1|n-1}], \mathbf{f}^H[\mathbf{\widehat{x}}_{i,n-1|n-1}]\Big]^T \label{augpex_est}\\
  \mathbf{M}_{i,n|n-1}^a &= \mathbf{F}_{i,n-1}^a\mathbf{M}_{i,n-1|n-1}^a\mathbf{F}^{aH}_{i,n-1} + \mathbf{Q}_n^a \label{augpeMSE}\\
	\mathbf{G}_{i,n}^a &= \mathbf{M}_{i,n|n-1}^a\mathbf{\underline{H}}^{aH}_{i,n}\big(\mathbf{\underline{H}}_{i,n}^a\mathbf{M}_{i,n|n-1}^a\mathbf{\underline{H}}^{aH}_{i,n} + \mathbf{\underline{R}}_{i,n}^a\big)^{-1} \label{aug_egain}\\
	\mathbf{\widehat{\underline{x}}}_{i,n|n}^a &= \mathbf{\widehat{x}}_{i,n|n-1}^a + \mathbf{G}_{i,n}^a\big(\mathbf{\underline{y}}_{i,n}^a - \mathbf{\underline{h}}_{i}^a[\mathbf{\widehat{x}}_{i,n|n-1}]\big)\label{augex_est}\\
	\mathbf{M}_{i,n|n}^a &= (\mathbf{I} - \mathbf{G}_{i,n}^a\mathbf{\underline{H}}_{i,n}^a)\mathbf{M}_{i,n|n-1}^a\label{augeMSE}
\end{align}
\hspace{0.15cm}$-$ For every node $i$, compute the diffusion update as
\begin{align}
	\mathbf{\widehat{x}}_{i,n|n}^a = \sum_{k \in \mathcal{N}_i} c_{k,i}\mathbf{\widehat{\underline{x}}}_{k,n|n}^a \label{augdex}
\end{align}
\caption{Distributed Augmented Complex EKF}
\label{alg:D-ACEKF}
\end{algorithm}
%
%
Algorithm \ref{alg:D-ACEKF} summarises the proposed   distributed augmented complex extended Kalman filter, where each node $i$ shares its (nonlinear) observation model $\mathbf{h}_i[\cdot]$ with its neighbours.
\newline
{
{\bf Remark \#5:} The D-ACEKF algorithm in Algorithm \ref{alg:D-ACEKF} extends the Distributed Extended Kalman filter in \cite{Sayed_EKF_2010} by using the widely linear model, and caters for the second order statistical moments of the state and noise models, together with the correlations present between the nodal observation noises.}
%
%
%
%
\section{Distributed Widely Linear Frequency Estimation}
The proposed augmented state space models are particularly suited for frequency estimation in power grid, as due to system inertia, the frequency  can be assumed identical over the network of measurement nodes, while  unbalanced systems generate noncircular measurements \cite{ mandic2014patent, Xia_Adaptive_Frequency_Freq_2012_IEEE_SP_Mag}. {For a three phase system, the instantaneous voltages at a node $i$ are given by}
\begin{eqnarray}
	{ v_{a,i,n} } & {=}& { V_{a,n}\cos(\omega nT + \phi) + z_{a,i,n} }\nonumber\\
	 {v_{b,i,n}} & {=}& {V_{b,n}\cos(\omega nT + \phi - 2\pi/3 + \Delta_b) + z_{b,i,n}} \nonumber\\
	{v_{c,i,n}} & {=}& {V_{c,n}\cos(\omega nT + \phi + 2\pi/3 + \Delta_c) + z_{c,i,n} }
	\label{Eq:3_phase_voltages}
\end{eqnarray}
where $V_{a,n}$, $V_{b,n}$ and $V_{c,n}$ are the amplitudes of the three-phase voltages at time instant $n$, $\omega = 2\pi f$ the angular frequency, $f$ the system frequency, $T$ the sampling interval, and $\phi$ the phase of the fundamental component, while $ z_{a,i,n}$, $ z_{a,i,n}$ and $ z_{a,i,n}$ are zero-mean observation noise processes. {The terms $\Delta_b, \Delta_c$ are used to indicate the phase distortions relative to a balanced three-phase system.} The phase voltages in ({\ref{Eq:3_phase_voltages}}) are first mapped to the complex voltage of orthogonal $\alpha$ and  $\beta$ components using Clarke's $\alpha \beta$ transformation, to give {
\begin{eqnarray}				
	\begin{bmatrix} v_{\alpha,i,n} \\  v_{\beta,i,n}\end{bmatrix} = 	\sqrt{\frac{2}{3}}\begin{bmatrix}  1 & -\frac{1}{2} & -\frac{1}{2} \\ 0 & \frac{\sqrt{3}}{2} & -\frac{\sqrt{3}}{2} \end{bmatrix} \begin{bmatrix} v_{a,i,n} \\ v_{b,i,n} \\ v_{c,i,n}\end{bmatrix}.
	\label{Eq:Clarke_transform}
\end{eqnarray}
}
The $\alpha \beta$ voltage is then converted to a scalar complex signal $v_{i,n}=v_{\alpha,i,n} + jv_{\beta,i,n}$. For balanced systems, in which $V_{a,n}=V_{b,n}=V_{c,n}$, { and $\Delta_b = \Delta_c = 0$}, the variables 
\begin{eqnarray*}				
	v_{\alpha,i,n}&=& A_n\cos(\omega nT + \phi) + z_{\alpha,i,n} \\
	v_{\beta,i,n} &=& A_n\cos(\omega nT + \phi + \frac{\pi}{2}) + z_{\beta,i,n}
\end{eqnarray*}
where $A_n = \frac{\sqrt{6}}{2}V_{a,n}$. The noises mapped via the $\alpha \beta$ transform now become
\begin{eqnarray*}				
	z_{\alpha,i,n} &=& \sqrt{2/3}\Big(z_{a,i,n} - \frac{1}{2}z_{b,i,n} - \frac{1}{2}z_{c,i,n}\Big) \\	
	z_{\beta,i,n}  &=& \sqrt{2/3}\Big( \frac{\sqrt{3}}{2}z_{b,i,n} - \frac{\sqrt{3}}{2}z_{c,i,n}\Big)
\end{eqnarray*}
For balanced systems, this scalar complex model takes a recursive form, so that for every node\footnote{The usual assumption in this type of estimation is that for a sampling frequency $>> 50$Hz, we have $A_n\approx A_{n-1}$.} 
\begin{eqnarray}				
	v_{i,n} &=& v_{\alpha,i,n} + jv_{\beta,i,n} \nonumber\\
			    &=& A_ne^{j(\omega nT + \phi)} + z_{i,n} \nonumber\\
			    &=& v_{i,n-1}e^{j\omega T} + z_{i,n}
	\label{Eq:output_voltage2}
\end{eqnarray}
where $z_{i,n} = z_{\alpha,i,n} + j \, z_{\beta,i,n}$. 

The state space model for this system at a node $i$ is shown in \eqref{Eq:ss_state4} and  \eqref{Eq:ss_obs4}, where the state variables $x_{k}$ and $u_n$ are used to estimate the exponential $e^{j\omega T}$ and the observation $v_{i,n}$ respectively, while $\mathbf{w}_n$ and $z_{i,n}$ are the state and observation noises respectively. 
The system frequency is then derived from the state variable $x$ as follows:
\begin{eqnarray}				
	\hat{f}_n &=& \frac{1}{2\pi T}\arcsin\big(\Im(x_n)\big)
	\label{Eq:freq_estimate1}
\end{eqnarray}
where $\Im(\cdot)$ is the imaginary part of a complex quantity.
%
\begin{algorithm}[t]
\caption{A Strictly Linear State Space (StSp-SL)}
State equation:
\begin{subequations}\label{alg:SS1_L}
\begin{align}
	\begin{bmatrix}x_n \\ u_n\end{bmatrix}
	= \begin{bmatrix}x_{n-1} \\ u_{n-1}x_n	\end{bmatrix} + \mathbf{w}_{n-1}
	\label{Eq:ss_state4}
\end{align}			
Observation equation:			
\begin{align}	 
	v_{i,n} = \begin{bmatrix} 0 & 1 \end{bmatrix} \begin{bmatrix}x_n \\ u_n\end{bmatrix} + z_{i,n}
	\label{Eq:ss_obs4}
\end{align}
\end{subequations}
\end{algorithm}
%
%
Figure \ref{fig:AB_trajectory} illustrates the trajectory of the transformed voltage (a rotating vector -- phasor), indicating that for a balanced system, Clarke's voltage $v_{i,n}$ in \eqref{Eq:output_voltage2} has a circular trajectory. 
However, the model in \eqref{Eq:ss_state4}  and \eqref{Eq:ss_obs4} becomes inaccurate when the system is operating under unbalanced conditions, in which case, the voltage amplitudes $V_{a,n}$, $V_{b,n}$ and $V_{c,n}$ are no longer equal or if the condition $\Delta_b = \Delta_c = 0$ is not satisfied, and the system trajectory becomes noncircular (ellipse in Figure \ref{fig:AB_trajectory}). For unbalanced systems, therefore, the correct system model is widely linear, and is given by 
{ 
\begin{align}				\nonumber
	v_{i,n} = & {}	\sqrt{\frac{2}{3}}  \begin{bmatrix}   1 & j  \end{bmatrix}  	
	\begin{bmatrix}  1 & -\frac{1}{2} & -\frac{1}{2} \\ 0 & \frac{\sqrt{3}}{2} & -\frac{\sqrt{3}}{2} \end{bmatrix} \begin{bmatrix} v_{a,i,n} \\ v_{b,i,n} \\ v_{c,i,n}\end{bmatrix} \\ \label{eq:cmplx_voltage}
   = & {} \sqrt{\frac{2}{3}} \begin{bmatrix}   1 & e^{j2\pi/3}  & e^{-j2\pi/3}  \end{bmatrix}  	
 \begin{bmatrix} v_{a,i,n} \\ v_{b,i,n} \\ v_{c,i,n}\end{bmatrix}. 
\end{align}
For the compactness of notation, we set $\theta_n = \omega n T + \phi$. Using the relationship $\cos(y) = \frac{e^{jy} + e^{-jy}}{2}$, the three phase voltages at each node $i$ can be rewritten as
\begin{align}			\nonumber
 \begin{bmatrix} v_{a,i,n} \\ v_{b,i,n} \\ v_{c,i,n} \end{bmatrix} = {} &   
\frac{1}{2}
\begin{bmatrix} 
V_{a,n} \left( e^{j\theta_n} +e^{-j\theta_n} \right)
  \\ 
V_{b,n} \left(e^{j(\theta_n - 2\pi/3 + \Delta_b) }  + e^{-j(\theta_n - 2\pi/3 + \Delta_b) } \right)
 \\
V_{c, n} \left(  e^{j(\theta_n + 2\pi/3 + \Delta_c) }  + e^{-j(\theta_n + 2\pi/3 + \Delta_c) } \right )
\end{bmatrix}	   
  \\ 
& {}  
+ \begin{bmatrix} z_{a,i,n} \\ z_{b,i,n} \\ z_{c,i,n} \end{bmatrix}  \label{eq:polarform}
\end{align}
Substituting \eqref{eq:polarform} into \eqref{eq:cmplx_voltage} gives
\begin{align} 
v_{i,n} 
   = {} & A_{n}e^{j\theta_n}   + B_{n}e^{-j\theta_n} + z_{i,n}
\label{Eq:output_voltageWL1}
\end{align}
where 
\begin{align}	
\label{Eq:A_B_coefficents}
A_n = {} & \frac{\sqrt{6}}{6}\left(V_{a,n} + V_{b,n}e^{j\Delta_b} + V_{c,n}e^{j \Delta_c} \right)\\ \nonumber
B_n = {} & \frac{\sqrt{6}}{6} \left(V_{a,n} + V_{b,n}e^{-j(\Delta_b + 2\pi/3)} + V_{c,n}e^{-j( \Delta_c - 2 \pi/3)} \right). 
\end{align}
}
%
\begin{figure}[t]
  \centering  
  \subfloat{\includegraphics[scale=0.7]{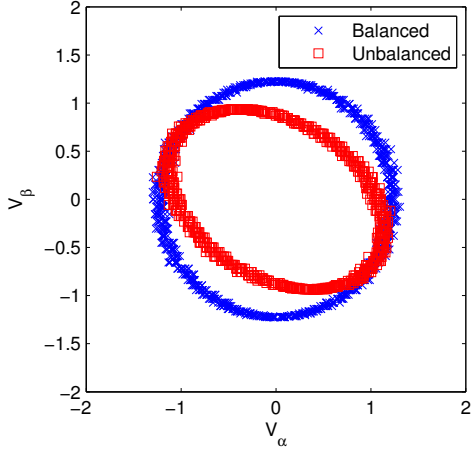}} \vspace{-0.35cm}
  \caption{Noncircularity in power systems. For a balanced system, characterised by $V_{a,n}=V_{b,n}=V_{c,n}$ and  $\Delta_b = \Delta_c = 0$, the trajectory of Clarke's voltage $v_{i,n}$ is circular (blue line). For unbalanced systems, the voltage trajectory is noncircular (red line), such as in the case of a 100\% single-phase voltage sag illustrated by the ellipse. In both cases the signal to noise ratio (SNR) was $20$dB.}
  \label{fig:AB_trajectory}
\end{figure}
%
For a balanced system under nominal conditions, we have $V_{a,n}=V_{b,n}=V_{c,n}$ and  $\Delta_b = \Delta_c = 0$, and the coefficient $B_n$ in ({\ref{Eq:A_B_coefficents}}) vanishes, so that the system is adequately characterised by the strictly linear model in \eqref{Eq:output_voltage2}. For  unbalanced systems, $B_n\neq0$ and the Clarke's voltage $v_{i,n}$ is noncircular, so that the expression in \eqref{Eq:output_voltageWL1} is second order optimal for both balanced and unbalanced conditions. This expression can be written in the form of a widely linear recursive model
\begin{eqnarray}
	v_{i,n} &=& v_{n-1}h_{n-1} + v^*_{n-1}g_{n-1}  + z_{i,n}
	\label{Eq:output_voltageWL2}
\end{eqnarray}
which is a first-order widely linear autoregressive model with coefficients $h_{k}$ and $g_{k}$. The widely linear (augmented) state space model corresponding to \eqref{Eq:output_voltageWL2} is defined in \eqref{Eq:ss_state3}, where the state vector consists of the strictly linear and conjugate weights $h_n$ and $g_n$, and the variable  $u_n$ which corresponds to the noise-free widely linear observation $v_n$. The system frequency can be computed from the state variables $h_n$ and $g_n$ as follows:
\begin{eqnarray}				
	\hat{f}_n &=& \frac{1}{2\pi T}\arcsin\big(\Im(h_n + a_ng_n)\big)
	\label{Eq:freq_estimate2}
\end{eqnarray}
%
%
\begin{eqnarray*}				
	a_n &=& \frac{-j\Im(h_n)+ j\sqrt{\Im^2(h_n) - |g_n|^2}}{g_n}
	\label{Eq:a1}
\end{eqnarray*}
The state space model in \eqref{Eq:ss_state3} and \eqref{Eq:ss_obs3} provides a realistic and robust characterisation of real world power systems, as  it represents both balanced and unbalanced systems, while its  nonlinear state equation also models the coupling between state variables. The StSp-WL model in \eqref{alg:SS1_WL}  can be implemented using the proposed distributed augmented complex extended Kalman filter in Section {\ref{Sec:D-ACEKF}}. 
%
%
\begin{algorithm}[t]
\caption{A Widely Linear State Space (StSp-WL)}
State equation:\vspace{-0.1cm}
\begin{subequations}\label{alg:SS1_WL}
\begin{align}
	\begin{bmatrix}h_n \\ g_n\\ u_n \\ h^*_n \\ g^*_n\\ u^*_n\end{bmatrix}
	 \!=\!  \begin{bmatrix}h_{n-1} \\ g_{n-1}\\ u_{n-1}h_{n-1} + u^*_{n-1}g_{n-1}\\ h^*_{n-1} \\ g^*_{n-1}\\ \!u^*_{n-1}h^*_{n-1} \!+\! u_{n-1}g^*_{n-1}\!\end{bmatrix}  \!+\! \mathbf{w}_{n-1}
	\label{Eq:ss_state3}
\end{align}			
Observation equation:	\vspace{-0.1cm}		
\begin{align}	 
	\begin{bmatrix} v_{i,n}\\ v^*_{i,n} \end{bmatrix} = \begin{bmatrix} 0 & 0 & 1 & 0 & 0 & 0 \\ 0 & 0 & 0 & 0 & 0 & 1\end{bmatrix} \begin{bmatrix}h_n \\ g_n\\ u_n \\ h^*_n \\ g^*_n\\ u^*_n\end{bmatrix} + \begin{bmatrix} z_{i,n}\\ z^*_{i,n} \end{bmatrix}
	\label{Eq:ss_obs3}
\end{align}
\end{subequations}
\end{algorithm}
%
\section{Frequency Estimation Examples}
\begin{figure}[t]
  \centering  
  \subfloat{ {\includegraphics[clip = true, trim = 25mm 40mm 270mm 15mm,scale=0.3]{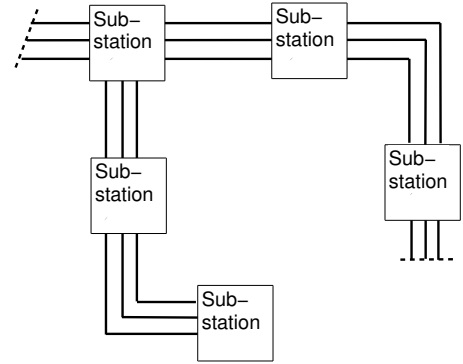}}}
  \caption{A distributed { power network} with $N=5$ nodes {(Sub-stations) }used in the simulations.}
  \label{fig:NetwotkTopologyN5}
\end{figure}
\begin{figure}[t]
  \centering  
  \subfloat{\includegraphics[scale=0.39]{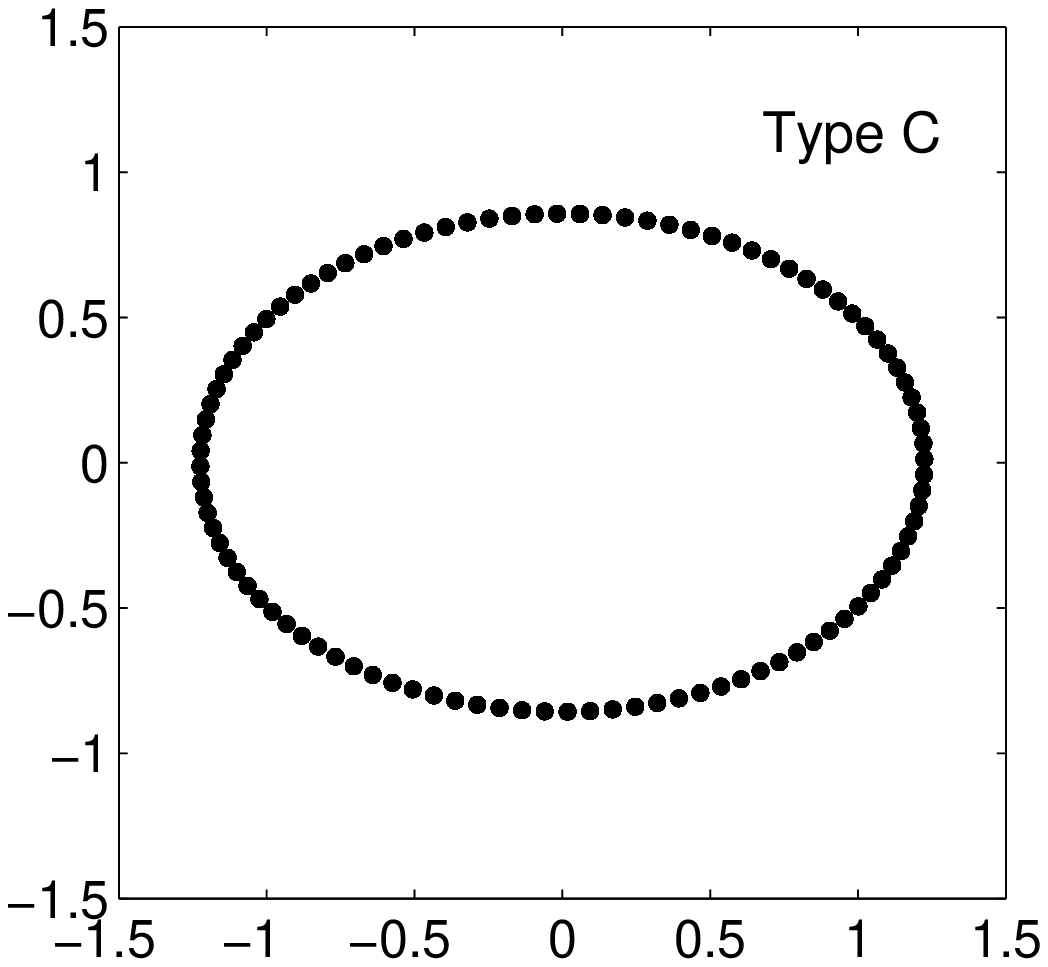}}
  \subfloat{\includegraphics[scale=0.39]{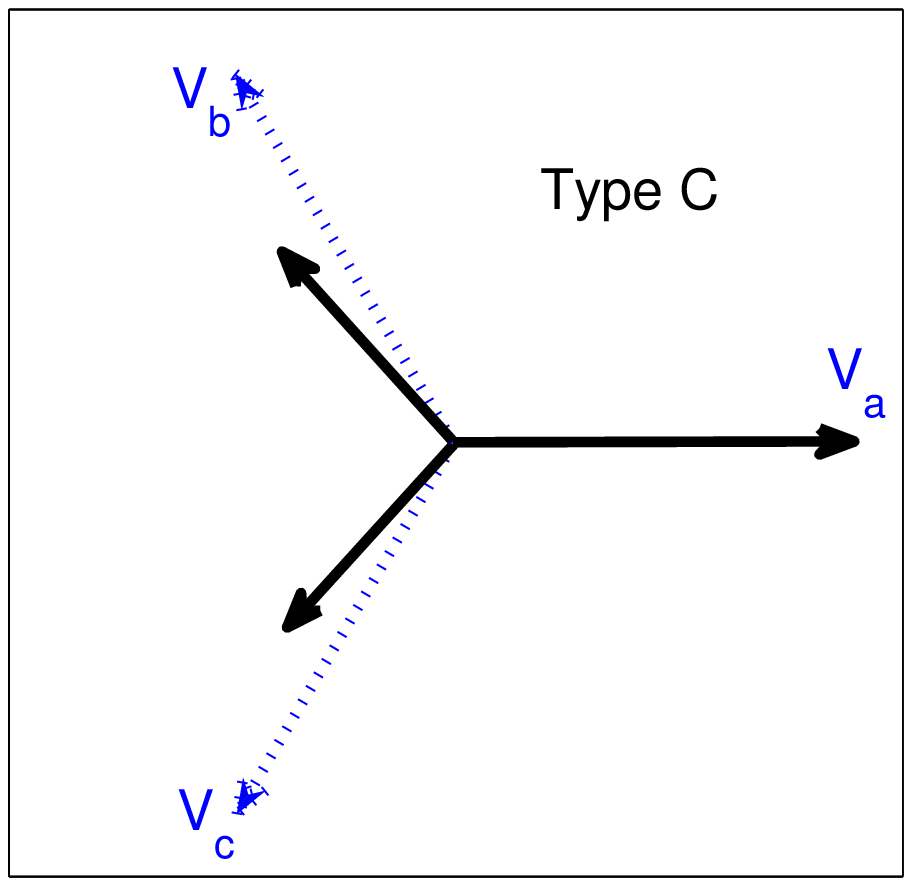}}\\ \vspace{-0.5cm}
  \subfloat{\includegraphics[scale=0.39]{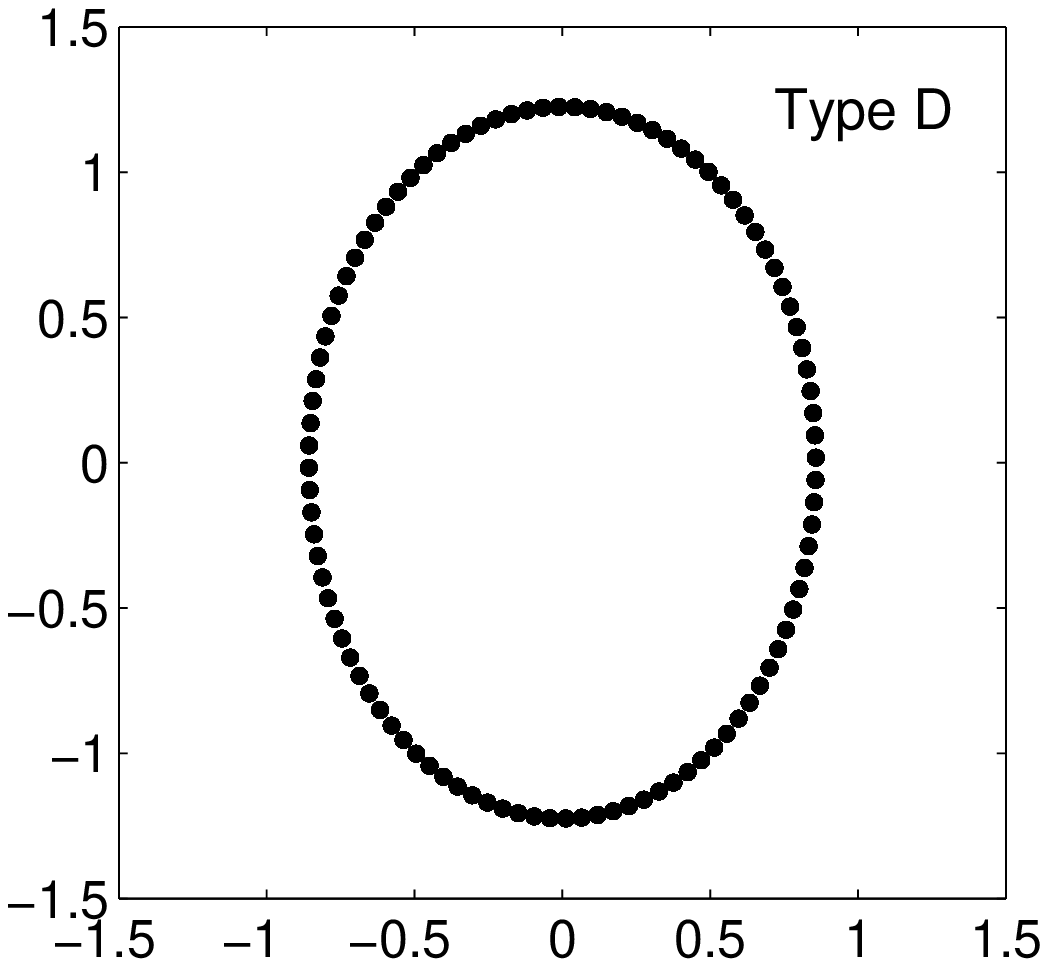}}
  \subfloat{\includegraphics[scale=0.39]{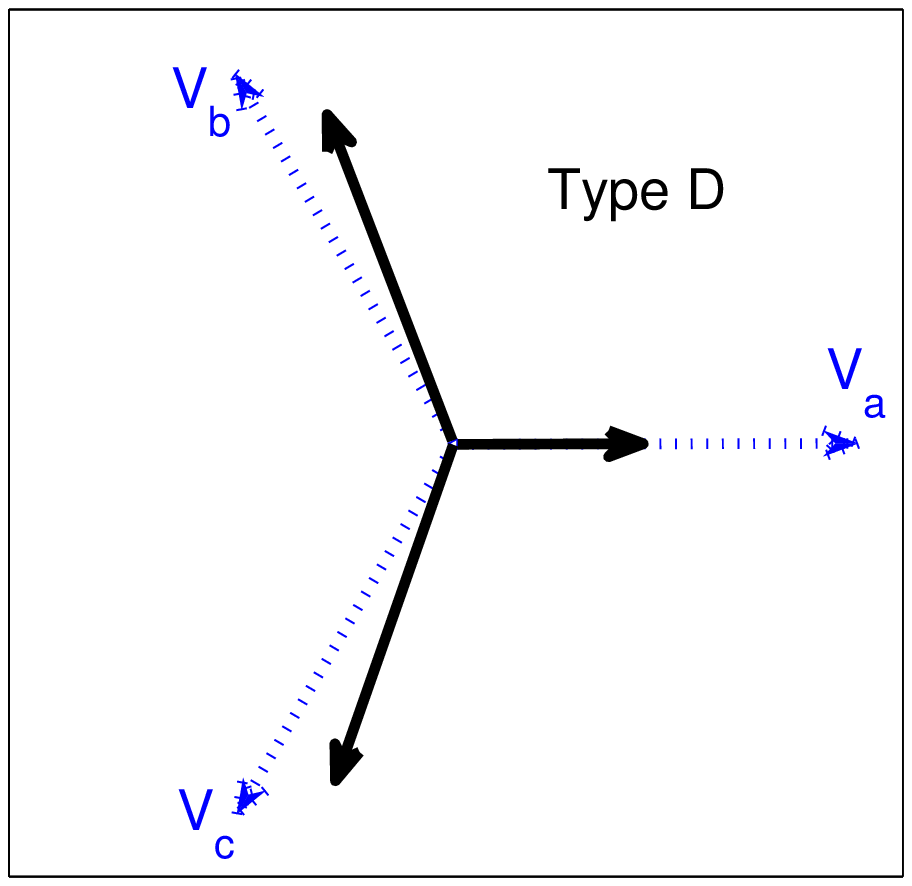}}  
  \caption{Geometric (left) and phasor (right) views of Type C and D unbalanced voltage sags. The real-imaginary phasor plots illustrate the noncircularity of Clarke's voltage in unbalanced conditions, indicated by the elliptical shapes of circularity plots. The parameters of this ellipse (degree of noncircularity) serve to identify the type of fault (in this case a voltage sag).}
  \label{fig:circularity_phasor_sags_C_D}
\end{figure}
\begin{figure*}[tbh]
  \centering  
  \subfloat{\includegraphics[scale=0.7]{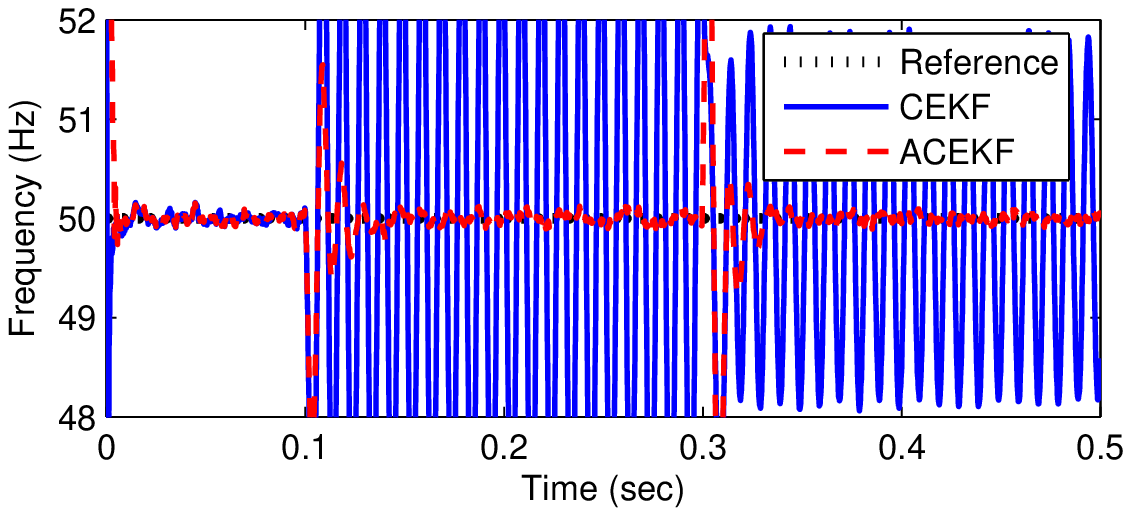}}
  \subfloat{\includegraphics[scale=0.7]{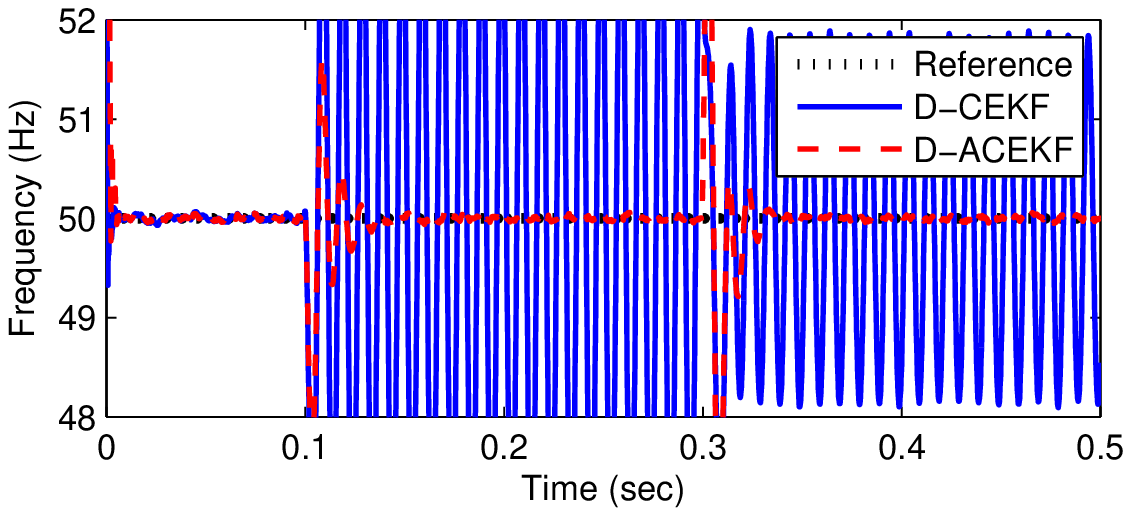}}
	\vspace{-0.35cm}
  \caption{Frequency estimation performance of single node (CEKF and ACEKF) and distributed (D-CEKF and D-ACEKF) algorithms for a system at $40$dB SNR. The system is balanced up to $0.1$s, it then undergoes a Type C voltage sag followed by a Type D voltage sag at 0.3s.}
  \label{fig:sag}\vspace{-0.5cm}
\end{figure*}
\begin{figure*}[tbh]
  \centering  
  \subfloat{\includegraphics[scale=0.7]{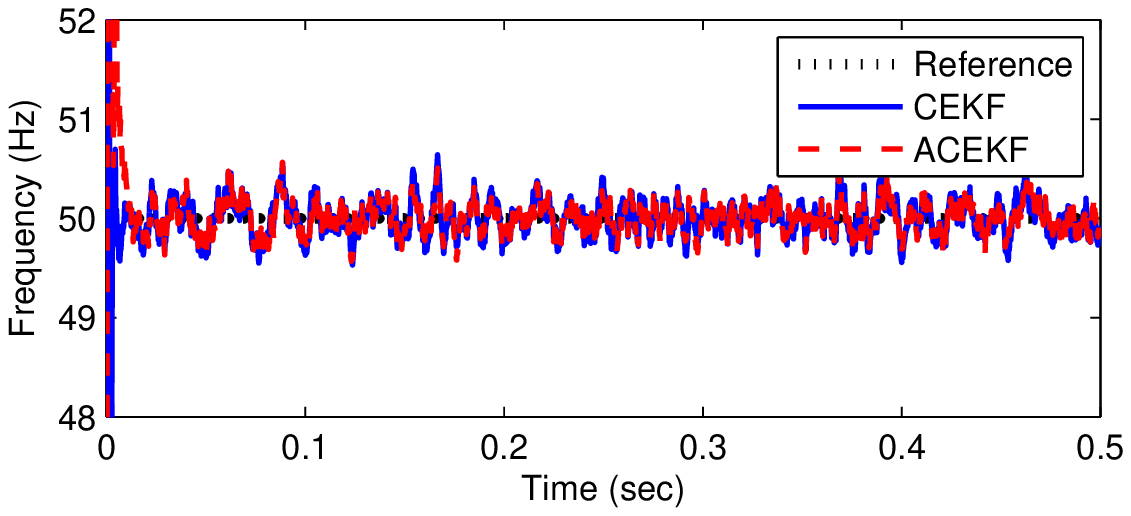}}
  \subfloat{\includegraphics[scale=0.7]{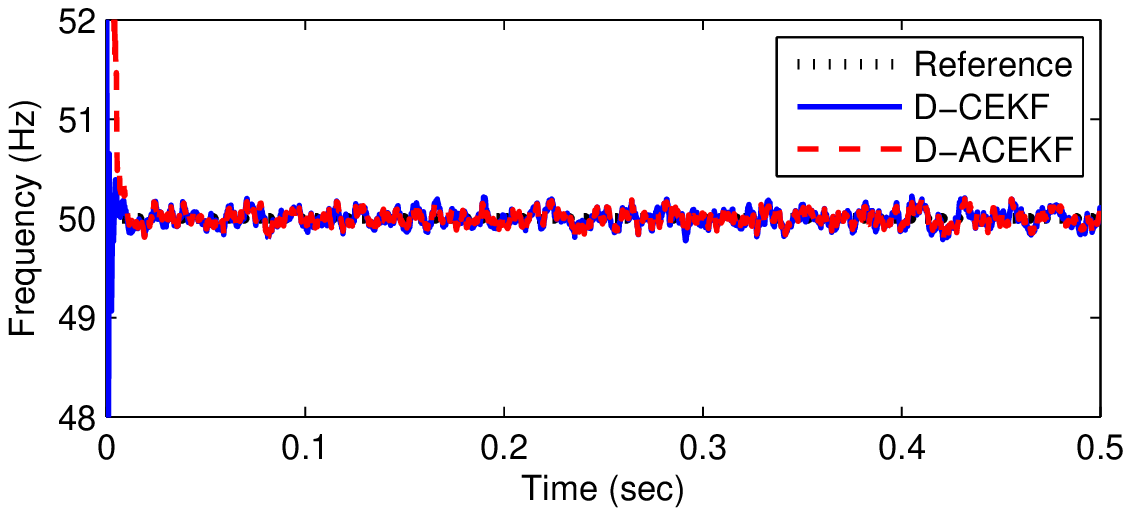}}
	\vspace{-0.35cm}
  \caption{Frequency estimation performance of single node (CEKF and ACEKF) and distributed (D-CEKF and D-ACEKF) algorithms for a balanced system in the presence of doubly white circular Gaussian noises at $30$dB SNR. As expected, the strictly and widely linear algorithms had similar performance.}
  \label{fig:freq_SNR30}\vspace{-0.5cm}
\end{figure*}
\begin{figure*}[tbh]
  \centering  
  \subfloat{\includegraphics[scale=0.7]{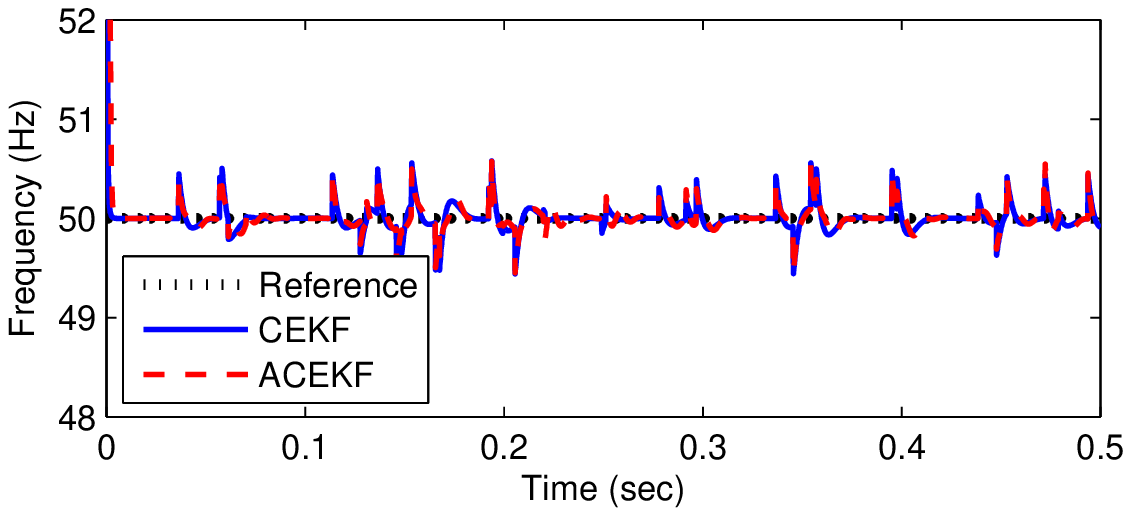}}
  \subfloat{\includegraphics[scale=0.7]{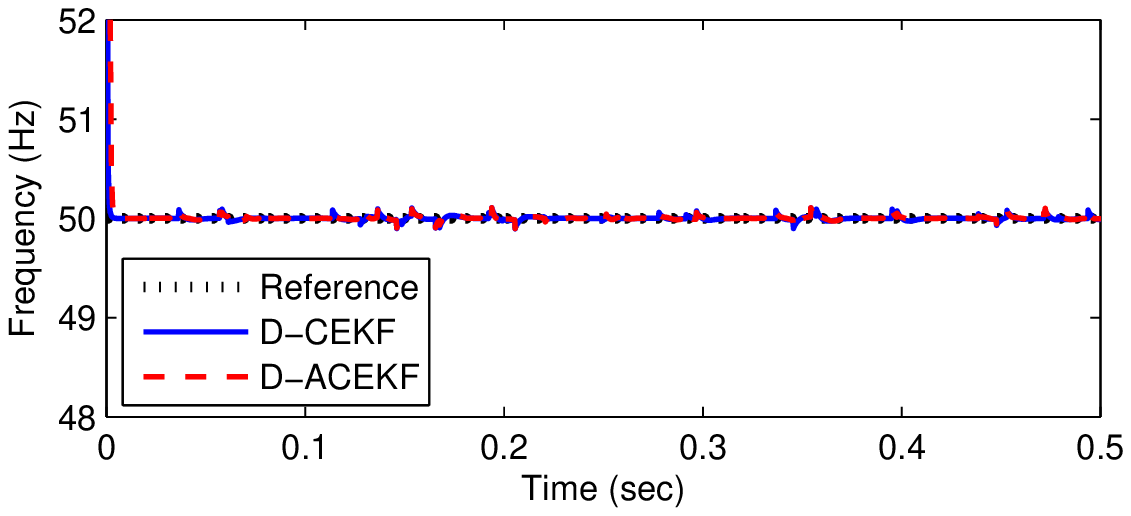}}
	\vspace{-0.35cm} 
  \caption{Frequency estimation performance of single node (CEKF and ACEKF) and distributed (D-CEKF and D-ACEKF) algorithms when the phase voltages at one of the nodes in the network are contaminated with random spike noise at $20$\% p.u.}
  \label{fig:spikeNoise}\vspace{-0.5cm}
\end{figure*}
\begin{figure*}[tbh]
  \centering  
  \subfloat{\includegraphics[scale=0.7]{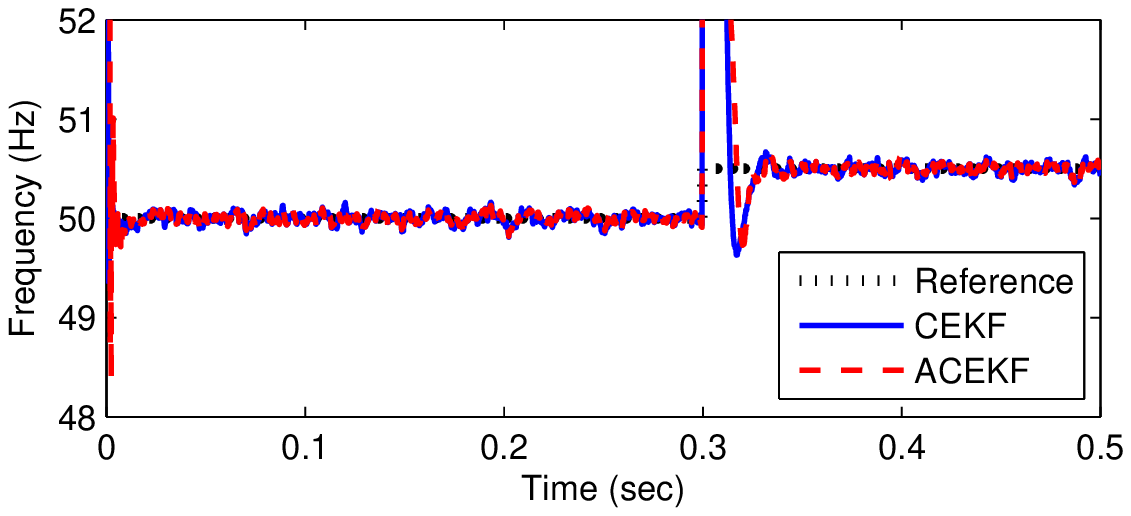}}
  \subfloat{\includegraphics[scale=0.7]{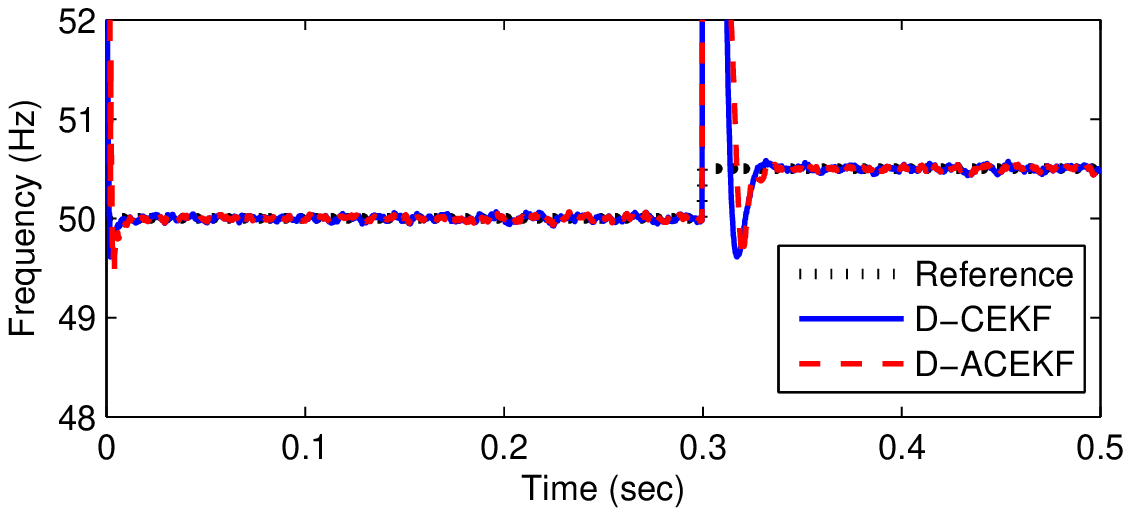}}
	\vspace{-0.35cm} 
  \caption{Frequency estimation performance of single node (CEKF and ACEKF) and distributed (D-CEKF and D-ACEKF) algorithms for a power system at $40$dB SNR, which experiences a step change in system frequency to $51$Hz.}
  \label{fig:stepChange}
\end{figure*}
\begin{figure*}[tbh]
  \centering  
 \subfloat {\includegraphics[clip = true, trim =0mm 0mm 0mm 10mm, scale=0.2]{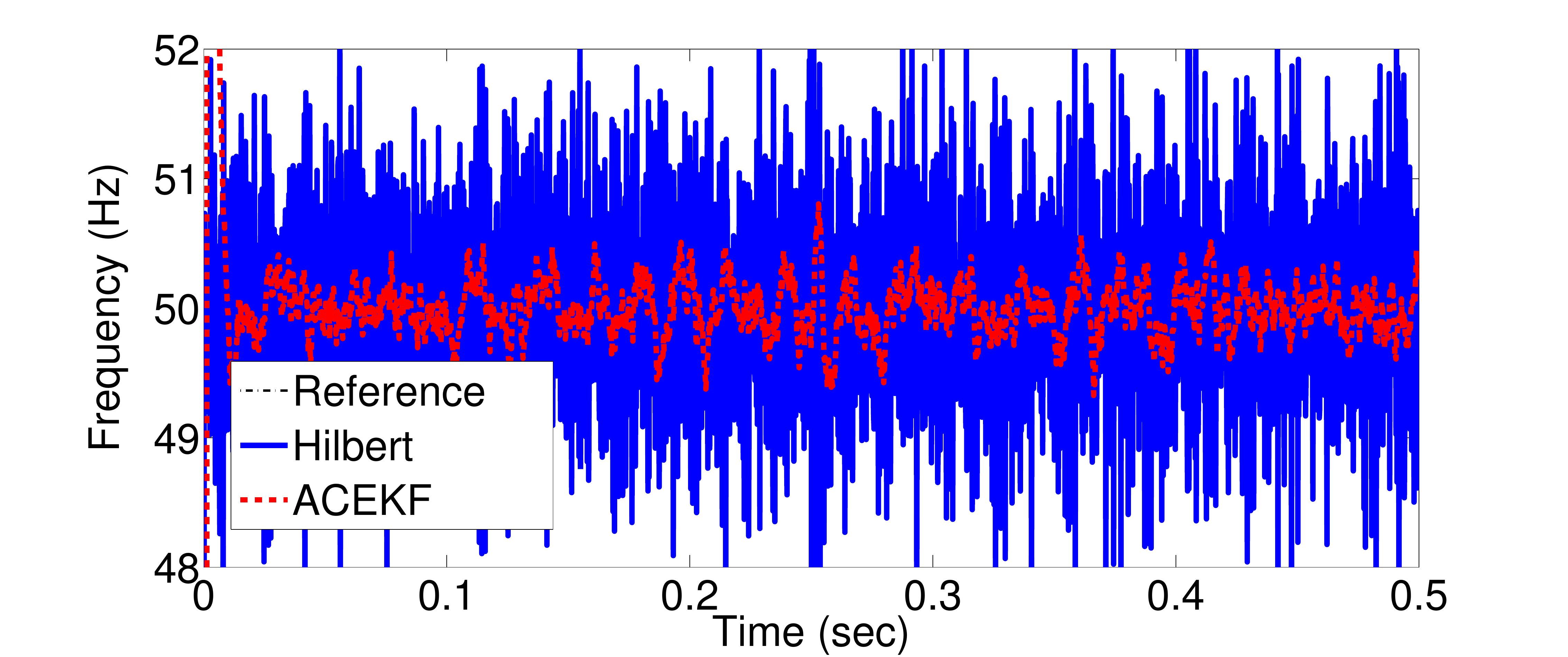}}
   \subfloat {\includegraphics[clip = true, trim =0mm 0mm 0mm 10mm, scale=0.2]{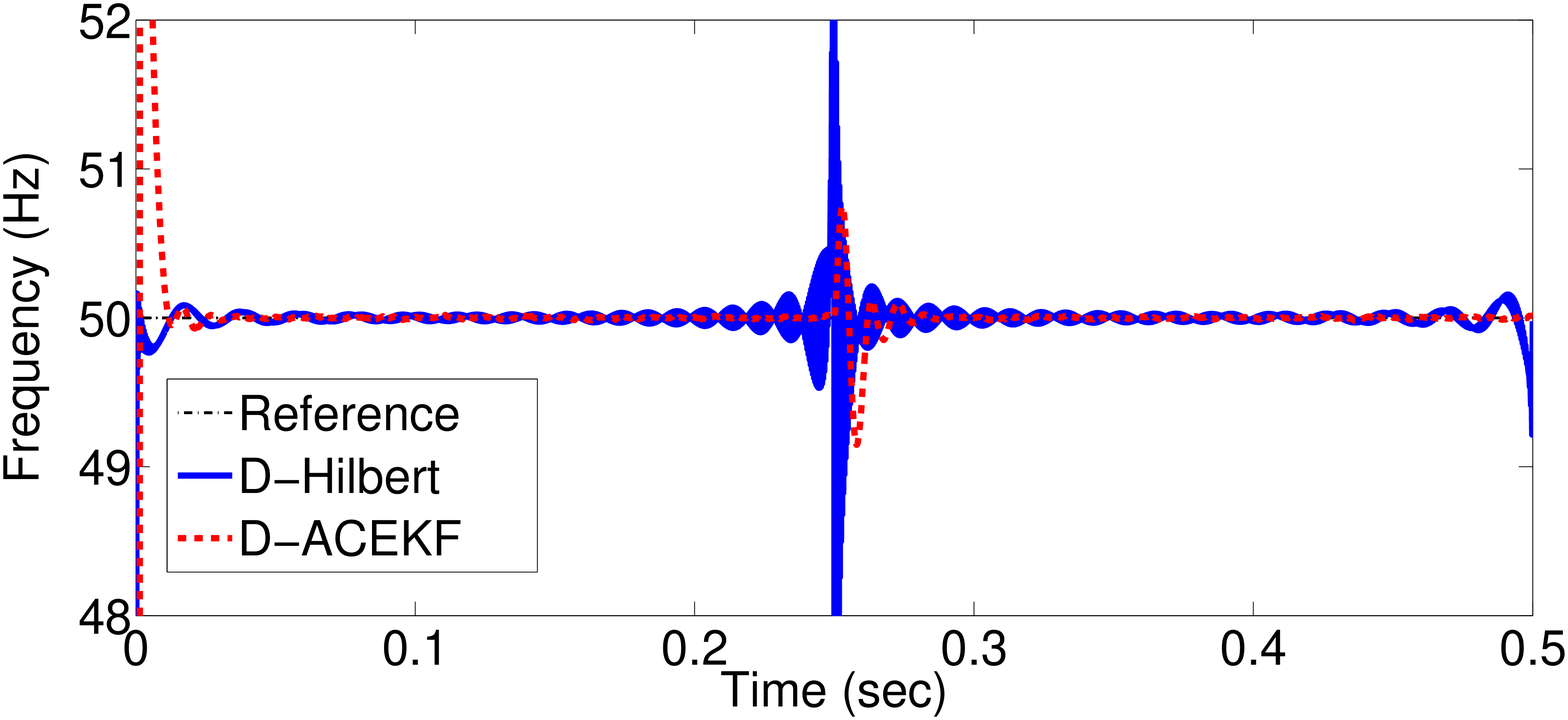}}
	\vspace{-0.35cm} 	
  \caption{{Frequency estimation performance comparison between Hilbert Transform based  instantaneous frequency estimate and ACEKF estimate. \textit{Left}: Single node based estimate (Hilbert and ACEKF). \textit{Right}: Distributed implementations of the algorithm (D-Hilbert and D-ACEKF) }}
  \label{fig:hilbert}\vspace{-0.5cm}
\end{figure*}
The performance of the algorithms was evaluated over both comprehensive illustrative simulation studies and a real world example. Within the synthetic data, the power system under consideration had a nominal frequency of $50$Hz, and was sampled at a rate of $5$kHz while the state vectors of all the nodes in the network were initialised to $50.5$Hz.  Without loss in generality, we used the distributed network topology shown in Figure \ref{fig:NetwotkTopologyN5}. The strictly linear state space model in  \eqref{alg:SS1_L}  and the widely linear state space model in  \eqref{alg:SS1_WL} were implemented using the proposed widely linear D-ACEKF and its strictly linear version D-CEKF. For rigour, the uncooperative CEKF and ACEKF were also considered.

\noindent \textbf{Case Study \#1: Voltage sags.} In the  first set of simulations, the performances of the algorithms were evaluated for an initially balanced system which became unbalanced after undergoing a Type C voltage sag starting at $0.1$s, characterised by a $20$\% voltage drop and $10^o$ phase offset on both the $v_b$ and $v_c$ channels, followed by a Type D sag starting at $0.3$s, characterised by a $20$\% voltage drop at line $v_a$ and a $10$\% voltage drop on both $v_b$ and $v_c$ with a $5^o$ phase angle offset. The degrees of noncircularity of these system imbalances are illustrated in Figure \ref{fig:circularity_phasor_sags_C_D}. Figure \ref{fig:sag} shows that, conforming with the analysis, the widely linear algorithms, ACEKF and D-ACEKF, were able to converge to the correct system frequency for both balanced and unbalanced operating conditions, while the  strictly linear algorithms, CEKF and D-CEKF, were unable to accurately estimate the frequency during the voltage sag due to under-modeling of the system (not accounting for its widely linear nature) -- see \eqref{Eq:output_voltageWL2}. As expected, the widely linear and strictly linear algorithms had similar performances under balanced conditions, as illustrated in the time interval $0$-$0.1$s. The distributed algorithms, D-CEKF and D-ACEKF, outperformed their uncooperative counterparts, CEKF and ACEKF, owing to the sharing of information between neighbouring nodes. 

\noindent \textbf{Case Study \#2: Presence of noise.} Figure \ref{fig:freq_SNR30} illustrates frequency estimation for a balanced system in the presence of white Gaussian noise at $30$dB SNR, while Figure \ref{fig:spikeNoise} illustrates frequency estimation in the presence of random spike noise, which typically occurs in the presence of switching devices or lightning.  The distributed algorithms, D-CEKF and D-ACEKF, outperformed their uncooperative counterparts, CEKF and ACEKF, because neighbouring nodes were able to share information to facilitate better estimation performances. In both cases, the distributed algorithms exhibited lower variance in the frequency estimate. 

\noindent \textbf{Case Study \#3: Frequency jumps.} Figure \ref{fig:stepChange} illustrates the performance of both single node and distributed algorithms (strictly and widely linear) when a power system is contaminated with white noise at $40$dB SNR and undergoes a step change in system frequency, a typical scenario  when generation does not match the load (microgrids and islanding). Although all the algorithms had similar responses to the step change in frequency, the distributed algorithms exhibited enhanced frequency tracking.
\begin{figure*}[ht!]
  \centering  
  \subfloat{\includegraphics[scale=0.7]{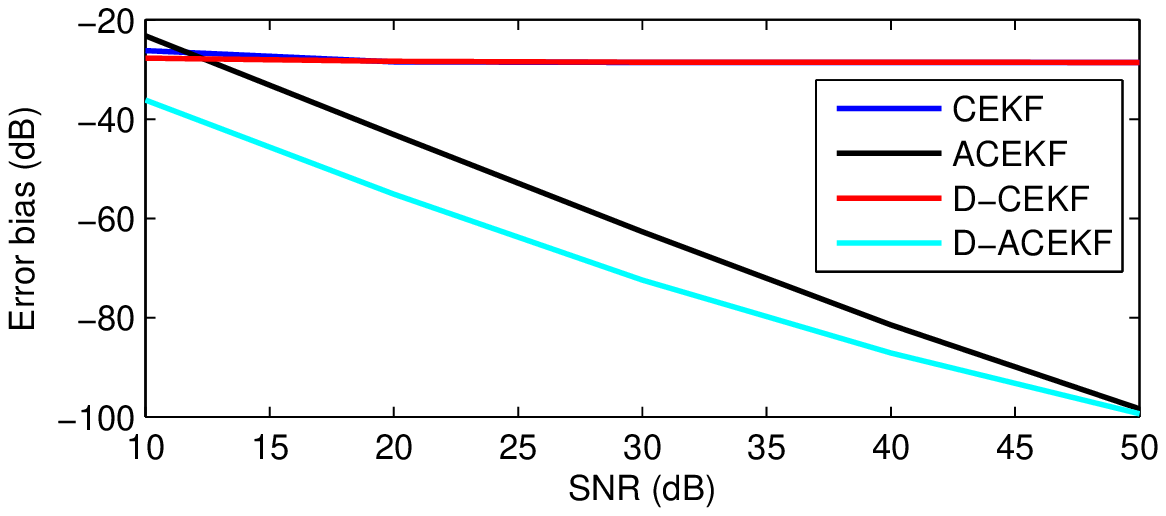}}
  \subfloat{\includegraphics[scale=0.7]{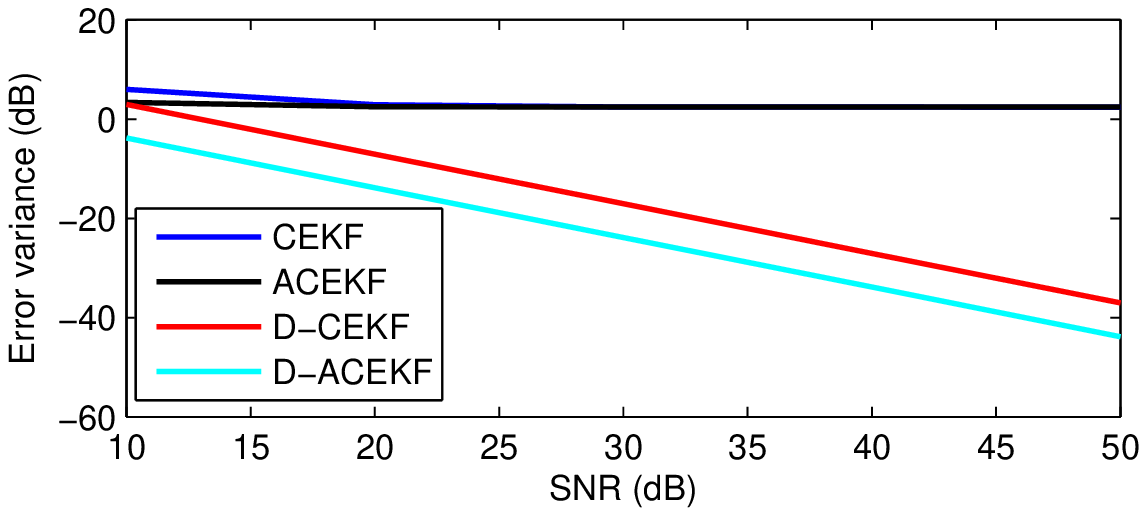}}
	\vspace{-0.35cm} 
  \caption{Bias and variance analysis of the proposed distributed state space frequency estimators  for an unbalanced system undergoing a type D voltage sag. {\it Left:} Estimation bias. {\it Right:} Estimation  variance.}
  \label{fig:BiasVar}
\end{figure*}

{
\noindent \textbf{Case Study \#4: Comparison with Hilbert method.} To further illustrate the advantages of widely linear modelling in a three-phase setting, the next case study compares the performance of the widely linear ACEKF to the Hilbert Transform based instantaneous frequency estimate from one of the three-phases. The three phase voltage signal at each node was simulated with circular white noise at  $30$dB SNR and Type D imbalance after $0.25$s.  Figure \ref{fig:hilbert} shows that the ``Hilbert" frequency estimate, found by differentiating the instantaneous phase angle of the single phase voltage that underwent the Hilbert Transform, was poor since the differentiation step (high pass filter) in the Hilbert method is not robust to noise.  This was observed both in single-node and distributed settings.
}

\noindent \textbf{Case Study \#5: Bias and variance of the proposed estimators.}  For rigour, Figure \ref{fig:BiasVar} provides the analysis of the bias and variance for the proposed distributed frequency estimators. The algorithms were evaluated at different SNR levels for an unbalanced system undergoing a Type D voltage sag (see also Figure {\ref{fig:circularity_phasor_sags_C_D}}). Observe that both  the single- and multiple-node widely linear algorithms, ACEKF and D-ACEKF, were asymptotically unbiased (left  panel, see Remark \#4) while both the single- and multiple-node strictly linear algorithms were biased.  In terms of the variance of the estimators (right panel), both the distributed estimation algorithms   outperformed their non-cooperative counterparts, while the only consistent estimator was the proposed   distributed augmented complex extended Kalman filter.

{
\noindent \textbf{Real World Case Study.} We next assessed the performance of the proposed algorithms on a real world case study using three-phase voltage measurements from two adjacent sub-stations in Malaysia\footnote{ {Due to data provenance issues we are unable to reveal which particular sub-stations the measurements originate from.}}  during a brief line-to-earth fault on the 29th June 2014. This caused voltage sags, similar to those in Case Study \#1. The three-phase measurements were sampled at 5kHz and the voltage values were normalized. The left panel in Figure \ref{fig:realWorld} shows the normalized $\alpha  \beta$ voltages at one of the sub-stations. The fault that occurs in phase A around 0.1s is reflected in the voltage dip in $v_{\alpha}$. The right panel in Figure \ref{fig:realWorld} shows the frequency estimate from the D-ACEKF and D-CEKF. Conforming with the analysis and the single node scenario in Figure \ref{fig:sag},  the collaborative widely linear D-ACEKF was able to track the real world frequency of a power network under both balanced and unbalanced conditions, whereas the strictly linear D-CEKF was unable to track the frequency after 0.1s when the line-to-earth fault (non-circularity) occurred. 

\begin{figure*}[tbh]
  \centering  
  \subfloat {\includegraphics[clip = true, trim =0mm 0mm 0mm 10mm, scale=0.2]{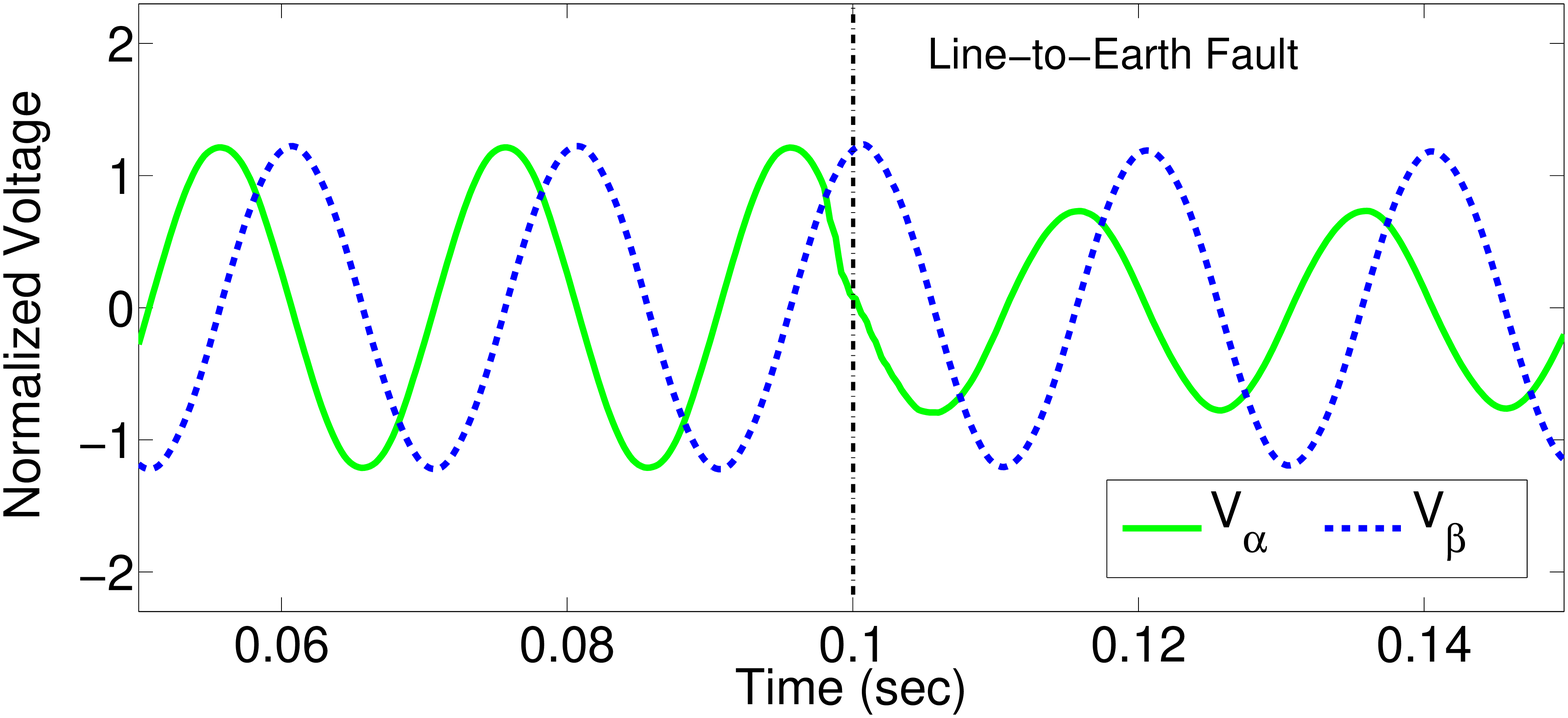}}
   \subfloat {\includegraphics[clip = true, trim =0mm 0mm 0mm 10mm, scale=0.2]{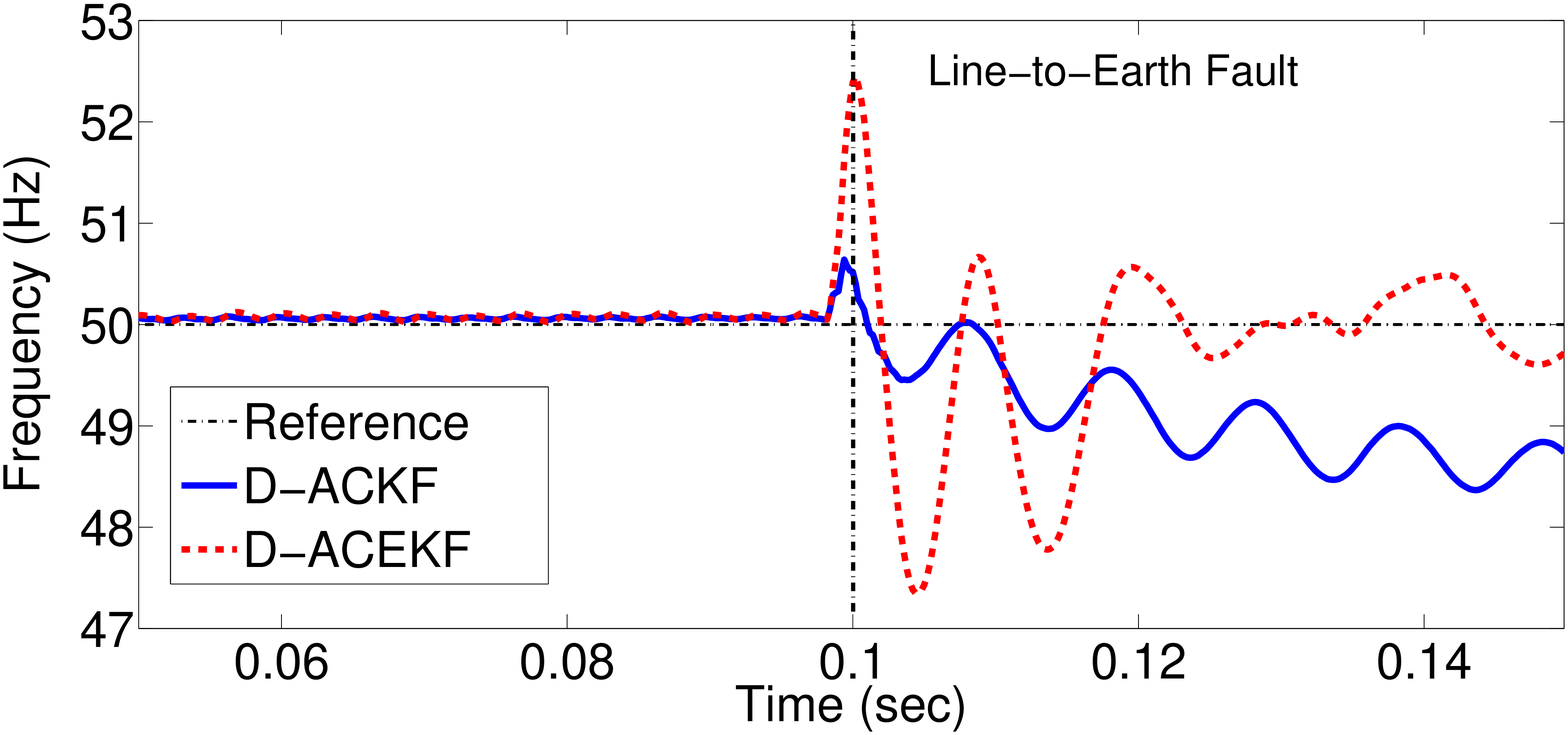}}
	\vspace{-0.35cm} 
  \caption{Real world case study. \textit{Left}: The $\alpha \beta$ voltages at Sub-station 1 before and during the fault event.       \textit{Right}: Frequency estimation using  the proposed algorithms before and during the fault event.}
  \label{fig:realWorld}
\end{figure*}

}

%
\section{Conclusions}

We have proposed a novel class of diffusion based distributed complex valued Kalman filters for cooperative frequency estimation in power systems. To cater for the general case of improper states, observations, and state and observation noises, we have introduced the  distributed (widely linear) augmented complex Kalman filter (D-ACKF) and its nonlinear version, the distributed augmented complex Kalman filter (D-ACEKF). These have been shown to provide sequential state estimation of the generality of complex signals, both circular and noncircular, within a general and unifying framework which also caters for correlated nodal observation noises.  
This novel widely linear framework has been applied for distributed state space based frequency estimation in the context of three-phase power systems, and has been shown to be optimal for both balanced and unbalanced operating conditions.  Simulations over a range of both balanced and unbalanced power system conditions {for both synthetic and real world measurements} have illustrated that the proposed distributed state space algorithms are consistent estimators, offering unbiased and minimum frequency estimation in both balanced an unbalanced system conditions, together with simultaneous frequency estimation and fault identification. 
%
%

\bibliographystyle{ieeetr} 
\balance
\bibliography{bibliography}

\begin{thebibliography}{10}

\bibitem{Bollen_DSP_Power_IEEE_SPM_2009}
M.~H.~J. Bollen, I.~Y.~H. Gu, S.~Santoso, M.~F. McGranaghan, P.~A. Crossley,
  M.~V. Ribeiro, and P.~F. Ribeiro, ``Bridging the gap between signal and
  power,'' {\em IEEE Signal Processing Magazine}, vol.~26, no.~4, pp.~11--31,
  2009.

\bibitem{Bollen_Book}
M.~H.~J. Bollen, {\em Understanding Power Quality Problems - Voltages and
  Interruptions}.
\newblock New York: IEEE Press, 1999.

\bibitem{Power_Standard}
``Power quality measurement methods, {IEC} 6100-4-15, 2003..''

\bibitem{Wang_DFT_2004}
{M. Wang} and {Y. Sun}, ``A practical, precise method for frequency tracking
  and phasor estimation,'' {\em IEEE Transactions on Power Delivery}, vol.~19,
  pp.~1547 -- 1552, Oct. 2004.

\bibitem{Lobos_Freq_DFT_1997}
{T. Lobos} and {J. Rezmer}, ``Real-time determination of power system
  frequency,'' {\em IEEE Transactions on Instrumentation and Measurement},
  vol.~46, pp.~877 --881, Aug 1997.

\bibitem{Xiao_Freq_est_LMS_IEEE_Trans_2005}
{Y. Xiao}, {R. K. Ward}, {L. Ma}, and {A. Ikuta}, ``A new {LMS}-based {F}ourier
  analyzer in the presence of frequency mismatch and applications,'' {\em IEEE
  Transactions on Circuits and Systems}, vol.~52, pp.~230 -- 245, Jan. 2005.

\bibitem{Huang_Freq_est_CEKF_IEEE_Trans_2008}
{C. H. Huang}, {C. H. Lee}, {K. J. Shih}, and {Y. J. Wang}, ``Frequency
  estimation of distorted power system signals using a robust algorithm,'' {\em
  IEEE Transactions on Power Delivery,}, vol.~23, pp.~41 --51, Jan. 2008.

\bibitem{Dash_freq_est_1999}
{P. K. Dash}, {A. K. Pradhan}, and {G. Panda}, ``Frequency estimation of
  distorted power system signals using extended complex {K}alman filter,'' {\em
  IEEE Transactions on Power Delivery}, vol.~14, pp.~761 --766, July 1999.

\bibitem{Hu_Kuh_State_MicroGrids_IJCNN_2011}
{Ying Hu}, {A. Kuh}, {A. Kavcic}, and {D. Nakafuji}, ``Real-time state
  estimation on micro-grids,'' in {\em The 2011 International Joint Conference
  on Neural Networks (IJCNN)}, pp.~1378--1385, 2011.

\bibitem{Mangesius_Obradovic_power_grid_2012}
{H. Mangesius}, {S. Hirche}, and {D. Obradovic}, ``Quasi-stationarity of
  electric power grid dynamics based on a spatially embedded {K}uramoto
  model,'' in {\em American Control Conference (ACC)}, pp.~2159--2164, 2012.

\bibitem{Clarke_Inst_Currents_Voltages_1951}
W.~C. Duesterhoeft, M.~W. Schulz, and E.~Clarke, ``Determination of
  instantaneous currents and voltages by means of alpha, beta, and zero
  components,'' {\em Transactions of the American Institute of Electrical
  Engineers}, vol.~70, no.~2, pp.~1248--1255, 1951.

\bibitem{Paap_Symmetrical_Components_2000}
G.~C. Paap, ``Symmetrical components in the time domain and their application
  to power network calculations,'' {\em IEEE Transactions on Power Systems},
  vol.~15, no.~2, pp.~522--528, 2000.

\bibitem{Clarke_Book}
E.~Clarke, {\em Circuit Analysis of A.C. Power Systems}.
\newblock New York: Wiley, 1943.

\bibitem{Beeman_1955}
{D. Beeman}, {\em Industrial Power System Handbook}.
\newblock McGraw-Hill, 1955.

\bibitem{Xia_Adaptive_Frequency_Freq_2012_IEEE_SP_Mag}
{{Y. Xia} and {S. C. Douglas} and {D. P. Mandic}}, ``Adaptive frequency
  estimation in smart grid applications: Exploiting noncircularity and widely
  linear adaptive estimators,'' {\em IEEE Signal Processing Magazine}, vol.~29,
  no.~5, pp.~44--54, 2012.

\bibitem{mandic2014patent}
D.~Mandic, Y.~Xia, and D.~Dini, ``{WO2014053610 - Frequency Estimation},''
  April 2014.
\newblock WO Patent App. PCT/EP2013/070,654.

\bibitem{Dini_Three_Phase_Freq_IEEE_IM_2013}
{D. H. Dini} and {D. P. Mandic}, ``Widely linear modeling for frequency
  estimation in unbalanced three{-}phase power systems,'' {\em IEEE
  Transactions onInstrumentation and Measurement}, vol.~62, no.~2,
  pp.~353--363, 2013.

\bibitem{Stadter_Dist_Spacecraft_IEEEMag_2002}
{P. A. Stadter}, {A. A. Chacos}, {R. J. Heins}, {G. T. Moore}, {E. A. Olsen},
  {M. S. Asher}, and {J. O. Bristow}, ``Confluence of navigation,
  communication, and control in distributed spacecraft systems,'' {\em IEEE
  Aerospace and Electronic Systems Magazine}, vol.~17, pp.~26 --32, May 2002.

\bibitem{Mandic_Fusion_Book_Short_2008}
D.~et~al., {\em Signal Processing Techniques for Knowledge Extraction and
  Information Fusion}.
\newblock Springer, 2008.

\bibitem{Cattivelli_Sayed_IEEETranAC_Dist_KF_2010}
{F. S. Cattivelli} and {A. H. Sayed}, ``Diffusion strategies for distributed
  {K}alman filtering and smoothing,'' {\em IEEE Transactions on Automatic
  Control}, vol.~55, pp.~2069 --2084, Sept. 2010.

\bibitem{Olfati_Saber_Flocking_2006}
R.~Olfati-Saber, ``Flocking for multi-agent dynamic systems{:} {A}lgorithms and
  theory,'' {\em IEEE Transactions on Automatic Control}, vol.~51, pp.~401 --
  420, March 2006.

\bibitem{Lopes_Sayed_TSP_Dist_LMS_2008}
{C. G. Lopes} and {A. H. Sayed}, ``Diffusion least-mean squares over adaptive
  networks: Formulation and performance analysis,'' {\em IEEE Transactions on
  Signal Processing}, vol.~56, pp.~3122 --3136, July 2008.

\bibitem{Yili_Dist_ACLMS_2011}
{Y. Xia}, {D. P. Mandic}, and {A. H. Sayed}, ``An adaptive diffusion augmented
  {CLMS} algorithm for distributed filtering of noncircular complex signals,''
  {\em IEEE Signal Processing Letters}, vol.~18, pp.~659 --662, Nov. 2011.

\bibitem{Olfati_Saber_Dist_KF}
{R. Olfati-Saber}, ``Distributed {K}alman filtering for sensor networks,'' in
  {\em 46th IEEE Conference on Decision and Control}, pp.~5492 --5498, Dec.
  2007.

\bibitem{Khan_Dist_KF_IEEETransSP_2008}
{U. A. Khan} and {J. M. F. Moura}, ``Distributing the {K}alman filter for
  large-scale systems,'' {\em IEEE Transactions on Signal Processing,},
  vol.~56, pp.~4919 --4935, Oct. 2008.

\bibitem{Dini_Class_WLKF_IEEE_TNNLS_2012}
{D. H. Dini} and {D. P. Mandic}, ``A class of widely linear complex {K}alman
  filters,'' {\em IEEE Transactions on Neural Networks and Learning Systems},
  vol.~23, pp.~775 --786, May 2012.

\bibitem{Gao_Dist_Cooperative_IEEETranComm_2011}
{Z. Gao}, H.-Q. Lai, and {K. J. R. Liu}, ``Differential space-time network
  coding for multi-source cooperative communications,'' {\em IEEE Transactions
  on Communications}, vol.~59, pp.~3146 --3157, Nov. 2011.

\bibitem{Mao_Wireless_Comm_IEEETranIFS_2007}
{Y. Mao} and {M. Wu}, ``Tracing malicious relays in cooperative wireless
  communications,'' {\em IEEE Transactions on Information Forensics and
  Security}, vol.~2, pp.~198 --212, June 2007.

\bibitem{Xia_WL_Freq_2011}
{{Y. Xia} and {D. P. Mandic}}, ``Widely linear adaptive frequency estimation of
  unbalanced three-phase power systems,'' {\em IEEE Transactions on
  Instrumentation and Measurement}, vol.~61, pp.~74 --83, Jan. 2012.

\bibitem{Picinbono97}
B.~Picinbono and P.~Bondon, ``{Second-order Statistics of Complex Signals},''
  {\em IEEE Transactions on Signal Processing}, vol.~45, no.~2, pp.~411--420,
  1997.

\bibitem{Moreno_2008}
{J. Navarro-Moreno}, ``{ARMA} prediction of widely linear systems by using the
  innovations algorithm,'' {\em IEEE Transactions on Signal Processing},
  vol.~56, pp.~3061 --3068, july 2008.

\bibitem{ErikssonAugmentID06}
J.~Eriksson and V.~Koivunen, ``{Complex Random Vectors and ICA Models:
  Identifiability, Uniqueness, and Separability},'' {\em IEEE Transactions on
  Information Theory}, vol.~52, no.~3, pp.~1017--1029, 2006.

\bibitem{Moreno_2009}
{J. Navarro-Moreno}, {J. Moreno-Kayser}, {R. M. Fernandez-Alcala}, and {J. C.
  Ruiz-Molina}, ``Widely linear estimation algorithms for second-order
  stationary signals,'' {\em IEEE Transactions on Signal Processing}, vol.~57,
  no.~12, pp.~4930--4935, 2009.

\bibitem{Hayes96}
M.~H. Hayes, {\em Statistical Digital Signal Processing and Modeling}.
\newblock John Wiley \& Sons, 1996.

\bibitem{Kar_Moura_Dist_KF_IEEETransSP_2011}
{S. Kar} and {J. M. F. Moura}, ``Gossip and distributed {K}alman filtering:
  Weak consensus under weak detectability,'' {\em IEEE Transactions on Signal
  Processing}, vol.~59, pp.~1766 --1784, April 2011.

\bibitem{Bru_Convergence_1994}
{R. Bru}, {L. Elsner}, and {M. Neumann}, ``Convergence of infinite products of
  matrices and inner-outer iteration schemes,'' {\em Electronic Transactions on
  Numerical Analysis}, vol.~2, pp.~183 --193, Dec. 1994.

\bibitem{Li_Optimal_Dist_fusion_2003}
{X.R. Li}, {Y. Zhu}, {J. Wang}, and {C. Han}, ``Optimal linear estimation
  fusion: Part {I}: Unified fusion rules,'' {\em IEEE Transactions on
  Information Theory}, vol.~49, pp.~2192 -- 2208, Sep. 2003.

\bibitem{Sayed_Proceedings_2014}
A.~Sayed, ``Adaptive networks,'' {\em Proceedings of the IEEE}, vol.~102,
  pp.~460--497, April 2014.

\bibitem{Sayed_arXiv}
A.~H. {Sayed}, ``{Diffusion Adaptation over Networks},'' {\em Academic Press
  Library in Signal Processing}, vol.~3, pp.~323--454, 2014.

\bibitem{Dini_Dist_ACKF_ASILOMAR_2012}
{D. H. Dini} and {D. P. Mandic}, ``Cooperative adaptive estimation of
  distributed noncircular complex signals,'' in {\em Conference Record of the
  Forty Sixth Asilomar Conference on Signals, Systems and Computers
  (ASILOMAR)}, pp.~1518--1522, 2012.

\bibitem{Dini_ACEKF_UDRC_2011}
{D. H. Dini} and {D. P. Mandic}, ``Widely linear complex extended {K}alman
  filters,'' in {\em Sensor Signal Processing for Defence Conference}, 2011.

\bibitem{Sayed_EKF_2010}
F.~Cattivelli and A.~Sayed, ``Distributed nonlinear {K}alman filtering with
  applications to wireless localization,'' in {\em IEEE International
  Conference on Acoustics Speech and Signal Processing (ICASSP)},
  pp.~3522--3525, March 2010.

\end{thebibliography}
\end{document}